\documentclass[reqno,11pt]{amsart}
\usepackage{graphicx}
\usepackage{amssymb}
\usepackage{amsmath}
\usepackage{amsfonts}
\usepackage{amsthm}
\usepackage{color}                   %Allows for colored text.
\usepackage{hyperref}
\usepackage[mathscr]{eucal}
\usepackage{enumerate}           
\usepackage{times}
\usepackage{authblk}
\usepackage[foot]{amsaddr}
\usepackage{ulem}                    %Strike out text passages with \sout{}.
\usepackage{enumitem}            %Modify indentation in itemize environment.

\numberwithin{equation}{section}
\allowdisplaybreaks[1]

\newtheorem{Def}{Definition}[section]
\newtheorem{Thm}[Def]{Theorem}

\newtheorem{Lemma}[Def]{Lemma}

\newtheorem{Corollary}[Def]{Corollary}

\newcommand{\beq}{\begin{equation}}
\newcommand{\eeq}{\end{equation}}
\newcommand{\Proof}{\begin{proof}}
\newcommand{\QED}{\end{proof} \noindent}

\newcommand{\mm}{\hspace{-.08cm}\cdot \hspace{-.08cm}}

\newcommand{\M}{\mathcal{M}}

\newcommand{\R}{\mathbb{R}}

\newcommand{\Gammati}{\tilde{\Gamma}}
\newcommand{\Riem}{{\rm Riem}}
\newcommand{\A}{\mathcal{A}}
\newcommand{\Amax}{\mathcal{A}^\text{max}}
\newcommand{\ess}{{\rm ess}}

\title[Regularizing Singularities]{The essential regularity of singular connections in Geometry}

\author[M.\ Reintjes]{Moritz Reintjes$^*$}
\address[*]{Department of Mathematics\\ City University of Hong Kong \\ SAR Hong Kong}
\email{moritzreintjes@gmail.com}

\author[B.\ Temple]{Blake Temple$^{**}$ \\ \\ December 30, 2025}
\address[**]{Department of Mathematics\\ University of California\\ Davis, CA 95616\\ USA}
 \email{temple@math.ucdavis.edu}

\begin{document} 

\begin{abstract} 
This paper, a culmination of the authors' theory of the RT-equations, accomplishes the following: (i) We discover there is a true (geometric) regularity associated with every affine connection, its {\it essential regularity}, the highest possible regularity achievable by coordinate transformation, a geometric property independent of starting atlas.  (ii) We give a checkable necessary and sufficient condition for determining whether or not a connection is at its essential regularity in a given atlas, based on the relative regularity of the connection and its Riemann curvature. (iii) We introduce a computable procedure based on the RT-equations for lifting any $L^p$ affine connection given in a starting atlas, to a new atlas in which the connection exhibits its essential regularity.  This resolves the long-standing problem of determining whether or not a singularity in an affine connection is removable or essential, applicable to any connection with components locally in $L^p$, $p>n$, general enough to include GR shock wave and cusp singularities in General Relativity. Since a manifold by itself does not carry an intrinsic level of regularity, the authors propose that the essential regularity of a connection marks the point at which an intrinsic level of regularity enters the subject of geometry.
\end{abstract}

\maketitle 

%\setcounter{tocdepth}{1}
%\small
%\tableofcontents
%\normalsize

%Keywords:  General Relativity; (Apparent) Singularities; Geometric Analysis; Elliptic Partial Differential Equations; RT-equations; Geometric Connection Regularity; Riemann Curvature; Shock Wave Theory.

\section{Introduction}  \label{Sec_intro}

The problem of determining whether or not a singularity is removable by coordinate transformation, and the true regularity of removable singularities, has been a central open problem in Differential Geometry and General Relativity from the beginning. In this paper we resolve the problem for any affine connection locally in $L^p$ in a given atlas, $p>n$. The question whether a singularity is essential or removable became central to General Relativity (GR) when Eddington and Finkelstein first discovered an explicit coordinate transformation which regularized the singularity present at the event horizon of the black hole in the Schwarzschild metric. Since that time, every reader is left wondering how one could determine at the start that the black hole singularity is removable. This question extends to other important singularities in GR, including singularities in the Kerr and Reissner-Nordstrom metrics, singularities at GR shock waves \cite{Israel, SmollerTemple, GroahTemple, ReintjesTemple_first, Reintjes_1, BarnesLeFlSchmSt}, as well as infinite gradient singularities like those that arise in numerical simulations of binary black hole mergers \cite{Alcubierre, AnninosCMSST, Pretorius}. In fact, the question how to tell, from a map alone, whether a singularity, like the North Pole in the Mercator map, is a problem with the map or a real singularity in the object which is being mapped, is as old as the subject of Differential Geometry.    Historically, determining that a singularity is removable has required the explicit construction of a regularizing coordinate transformation, and failing this, neither a definitive test, nor an explicit procedure for constructing regularizing coordinate transformations, has been available for determining {\it a priori} whether the search for a regularization is feasible or futile \cite{Choquet, HawkingEllis, Israel, KazdanDeTurck, ReintjesTemple_first, Uhlenbeck}.  This problem extends to all levels of regularity, and more fundamentally raises the following question: What is the true, highest possible, regularity a metric, or more generally a connection, can exhibit in an atlas of coordinate charts, and what defines and characterizes this true, highest possible, geometric level of regularity of a connection?

In this paper we accomplish three things: (i) We make precise the notion of the highest possible regularity of a connection, its {\it essential regularity}, and establish its consistency as a geometric property of any affine connection, independent of starting atlas, cf. Definition \ref{Def1}. (ii) We give a necessary and sufficient condition, based on the relative regularity of a connection and its Riemann curvature tensor, for when a connection has reached its essential regularity in a given atlas, cf. Theorem \ref{Cor1}. (iii) We introduce an iterative procedure based on the RT-equations, starting from any atlas, for constructing a new atlas in which the essential regularity of a connection is achieved, cf. Theorem \ref{Thm1}. We establish (i) - (iii) in the Sobolev spaces $W^{s,p}_{\rm loc}$, $s \geq 0$, $p>n$, for any affine connection whose components are locally $L^p$ regular in some starting atlas on a general $n$-dimensional manifold $\M$. Connection regularity $L^p$, $p>n$, includes all H\"older and Lipschitz continuous metrics in the case of metric connections, thus addressing GR shock wave solutions constructed by the Glimm scheme as well as cusp singularities (continuous metrics with infinite gradients), both obstacles to numerical simulation, but not yet the discontinuous and unbounded metrics associated with black hole singularities.\footnote{By Christoffel's formula a metric is always exactly one derivative more regular than its connection. So a metric connection in $L^p$ has a metric in $W^{1,p}$. When $p>n$ such metrics are continuous by Morrey's inequality, but can have infinite gradient; when $p\leq n$ metric components can be discontinuous and unbounded, like the singularity at the event horizon of the Schwarzschild metric.}       The ideas in this paper potentially extend to other regularity classes and geometries, including connections on vector bundles \cite{ReintjesTemple_ell5}.

To accomplish results (i) - (iii) in this paper, we prove that the {\it RT-equations}, originally designed by the authors to locally lift the regularity of singular connections by one derivative, surprisingly, also induce an implicit hidden regularization of the Riemann curvature tensor, together with a {\it global} regularization of transition maps between regularizing coordinate charts. This is a culmination of the authors' prior results on local one-step regularizations which began with the discovery of the RT-equations.

The RT-equations, introduced by the authors' in \cite{ReintjesTemple_ell1} and analysed in \cite{ReintjesTemple_ell2, ReintjesTemple_ell3, ReintjesTemple_ell4, ReintjesTemple_ell6}, are a non-invariant coordinate dependent system of partial differential equations (PDE) which is elliptic regardless of metric signature, and whose solutions provide the local coordinate transformations which lift an $L^p$ connection by one derivative of regularity when its Riemann curvature is also in $L^p$. This extends the optimal regularity results of Kazdan-DeTurck from Riemannian to non-Riemannian geometries. See also \cite{Anderson, ChenLeFloch} for results in Lorentzian geometry subject to additional assumptions. Several applications of the RT-equations have been established so far, including the problem of existence and uniqueness of geodesic curves at low regularity \cite{ReintjesTemple_geodesics}, (cf. \cite{LangeLytchakSaemann, SaemannSteinbauer, Steinbauer}), Penrose's Strong Cosmic Censorship Conjecture \cite{Reintjes_SCC}, (cf. \cite{Penrose}), Hawking's Singularity Theorem \cite{KunzingerReStGo}, (cf. \cite{Graf, GrafGrKuSt, KunzingerStVi}), and the first extension of Uhlenbeck compactness \cite{Uhlenbeck} to vector bundles over Lorentzian and non-Riemannian manifolds allowing for compact as well as non-compact Lie groups \cite{ReintjesTemple_ell5}, (cf. \cite{ChenGiron, ReintjesTemple_ell6, Wehrheim}).  The authors' research program began with the problem of singular GR shock waves, constructed in coordinates where Lorentzian metrics are below their essential regularity \cite{BarnesLeFlSchmSt, GroahTemple, Israel, ReintjesTemple_first, ReintjesTemple_geo, SmollerTemple}. 

In our prior results we viewed the curvature regularity as fixed, which was sufficient for a one-step regularization by the RT-equations, but not for the multi-step regularization required to extend our theory to a consistent definition of the essential regularity of a connection and the coordinate transformations which achieve it, because multi-step regularizations always require a regularization of the curvature as well. The RT-equations do not provide any such explicit regularization of the curvature, and for this reason authors thought for a long time that multi-step regularizations by the RT-equations are not possible. But, in this paper, by leveraging the notion of essential regularity--a consistent notion of regularity not formulated before--we prove that an implicit (hidden) regularization of the curvature by the RT-equations takes place in each step until the essential regularity of the connection is reached. Moreover, our prior results were only sufficient to obtain a regularization in a neighborhood of any point, but this was not sufficient to regularize a geometry globally, because even when covering a manifold by regularizing coordinate charts, it seemed impossible to control the regularity of the transition maps. This problem is resolved in the present paper by discovering a correlation between Jacobian regularity and the regularity of connections a Jacobian maps between.\footnote{Myers-Steenrod \cite{MyersSteenrod}, Calabi-Hartman \cite{CalabiHartman} and Taylor \cite{Taylor_isometries} used similar ideas to control Jacobian regularity of transformations that preserve regularity of Riemannian metrics. McCann et al. extended this to Lorentzian metrics in \cite{McCann_etal}. We refine this result to a necessary and sufficient condition and to transformations between affine connections of {\it different} regularities (cf. Lemma 2.1), and this refinement is crucial for the proofs in this paper.}      Using this regularity control of Jacobians, together with a consistent notion of essential regularity and a refined existence theory for the RT-equations, we prove that the RT-equations regularize both a connection and its curvature by one derivative, iteratively, until the connection reaches its essential regularity. Moreover, we prove that connections regularized up to their essential regularity, starting from different coordinate charts, always generate transition maps which exhibit the requisite regularity required to preserve the essential regularity, thereby extending {\it essential regularity} to a consistent definition applicable to any global atlas in which connection components lie locally in $L^p$, $p>n$. Taken together, this establishes for any connection given in an atlas the existence of a new atlas in which the connection exhibits its essential regularity, thus achieving a global regularization of an entire geometry to its essential regularity, applicable to any $L^p$ connection defined in any starting atlas of regularity $W^{2,p}$ or higher, $p>n$. 

Turning to the problem of regularizing singularities, to clarify the ideas, consider first the simplest case of point singularities. The essential regularity of a connection at a point is the highest possible regularity one can obtain in some coordinate chart containing that point (cf. Definition \ref{Def_ess-reg}); and by a point {\it singularity} we mean a connection which is below its essential regularity in every restriction of some given coordinate chart to a neighborhood of the point singularity. In this case, the results in this paper imply that a necessary and sufficient condition for a connection $\Gamma$ given in a coordinate chart to be regularizable around a point, (i.e., for a point singularity to be removable), is that the Riemann curvature tensor $\Riem(\Gamma)= {\rm Curl}(\Gamma)+[\Gamma,\Gamma]$ be at least as regular as the connection in some restriction of the chart to some neighborhood of the point, due to a cancellation of singular terms in the curvature; and given the singularity is removable, solving the RT-equations successively until the curvature is one derivative less regular than the connection, provides an explicit iterative procedure for constructing the coordinate transformations which lift $\Gamma$ to its essential (highest possible) regularity in some new coordinate chart containing the point.\footnote{One can make an analogy here with the classical problem of determining whether or not a vector field $F$ is conservative: Given the components of a vector field $F$ in some coordinate system, a necessary and sufficient condition for there to exist a potential in a simply connected domain, is the condition $Curl F \equiv 0$;  and given $F$ is appropriately Curl-free, the method of partial integration provides a procedure for explicitly computing its potential.}   The authors' prior work addressed local one-step regularizations of point singularities, but singularities are generally global, that is, points or surfaces of regularity defects could be scattered, potentially densely, throughout the manifold, like Glimm scheme based GR shock waves. Based on our new global multi-step regularization of singular connections and transition maps by the RT-equations, and the global notion of essential regularity established in Definition \ref{Def1}, we obtain the following complete characterization of local (point) and global singularities. 

\vspace{.2cm} \noindent 
{\it The hierarchy of apparent singularities:}
A connection is two or more derivatives below its essential regularity if and only if the curvature is at least one derivative more regular than the connection; a connection is precisely one derivative below its essential regularity if and only if the curvature has the exact same regularity as the connection; a connection is at its essential regularity if and only if the curvature is exactly one derivative less regular than the connection, (see Theorem \ref{Cor1} for the precise statement).\footnote{The proof requires the theory of the RT-equations for non-Riemannian connections.   In the case of positive definite Riemannian metric connections, the methods of Kazdan-DeTurck appear to suffice in place of the RT-equations, but present authors are not aware of a statement of this result in the literature for Riemannian metric connections.} \vspace{.2cm}

\noindent 
Thus, at levels of regularity low enough to be considered problematic, an apparent (removable) singularity is one where the curvature is at least as regular as the connection, and an essential (non-removable) singularity is one where the curvature is precisely one derivative less regular than the connection. Again, our current theory establishes this principle rigorously for regularity measured in the Sobolev spaces $W^{s,p}_{\rm loc}$, $s\geq 0$, $p>n$. Based on this hierarchy of singularities, we prove in Theorem \ref{Thm1} that any apparent singularity in a connection can be removed by regularizing the connection to its essential regularity by iterative use of the RT-equations, both locally and globally. To the authors' knowledge, this is the first definitive theory of apparent singularities applicable to Lorentzian geometry and general affine connections. 

Even though Sobolev norms in non-Riemannian geometry are in general not invariant, we establish here that the highest possible level of Sobolev regularity, the essential regularity, is invariant. In \cite{ReintjesTemple_Measurements}, the authors argue that the essential regularity of the gravitational metric, determined experimentally by large length-scale measurements, would encode information about the underlying physics on smaller length scales. Because a manifold by itself does not have enough structure to determine a geometric level of regularity, (every $C^1$ atlas is compatible with a $C^\infty$ atlas \cite{Hirsch}), the essential regularity of a connection marks the point at which a geometric level of regularity enters the subject of geometry.   

The outline of the paper is as follows: In Section \ref{Sec_Results} we state and discuss the main results of this paper, including a discussion of $C^\infty$ structures. In Section \ref{Sec_RT} we establish an improved existence theory for the RT-equations, as required for this paper. In Section \ref{Sec_Jacobian} we prove regularity control of coordinate transformations when connections are mapped between different levels of regularity. Combining the results of Sections \ref{Sec_RT} and \ref{Sec_Jacobian}, we develop in Section \ref{Sec_local} the local versions of the results stated in Section \ref{Sec_Results}. In Section \ref{Sec_global} we complete the theory by extending the local results to global ones, thereby completing the proofs of the theorems stated in Section \ref{Sec_Results}. In Section \ref{Sec_estimates} we establish Sobolev estimates for one- and two-step regularizations of connections on compact manifolds.

\section{Statement of results} \label{Sec_Results}

Regularizing the components of a connection by coordinate transformation requires measuring the regularity in some space of differentiability, like $C^k$, $H^s$, $C^{k,\alpha}$ or $W^{s,p}$.  In this paper we measure differentiability in the Sobolev spaces $W^{s,p}$, the space of functions with $s$ derivatives integrable in $L^p$ with standard Lebesgue measure using the Euclidean metric of the coordinate charts in $\R^n$. Differentiation is then taken with respect to partial derivatives in $\R^n$.  For example, given a function $f$ on an open set $\Omega \subset \R^n$, we write $f \in W^{s,p}(\Omega)$ if $f \in L^p(\Omega)$ and with partial derivatives $\partial_{x}^k f \in L^p(\Omega)$ for all $k=1,...,s$, \cite{Evans}.   We are interested in the regularity of singular connections on bounded sets, so with slight abuse of terminology we define a function on an unbounded domain in $\R^n$ to be in $L^p$ if its $L^p$ norm on every bounded subset is finite, (essentially, we use $L^p$ to mean $L^p_{\rm loc}$ and $W^{s,p}$ for $W^{s,p}_{\rm loc}$).  For clarity we restrict the theory to non-negative integer values $s\in \mathbb{N}_0$ (including $s=0$) and fixed values of $p$, assuming $n<p<\infty$.\footnote{The ideas here are more general, and analogous results should in principal carry over to other spaces amenable to elliptic regularity theory applied to the RT-equations, e.g., $W^{s,p}$ for continuous values of $s\geq 0$. The ideas here extend to connections on vector bundles by the results in \cite{ReintjesTemple_ell5}. The assumption $p>n$ and $s\in \mathbb{N}_0$,  (and $p>4$ if $n\leq 3$ and $s=0$), is required by the authors' current existence theory for the RT-equations in \cite{ReintjesTemple_ell4} on which the results in this paper are based. Note also that the case $p=\infty$ is a singular case of elliptic regularity theory, and when $\Gamma$ and $\Riem(\Gamma)$ are in $L^\infty$ the RT-equations give a regularization to $W^{1,p}$ for any $p<\infty$.}                In this paper we address affine connections defined on a manifold $\mathcal{M}$ with atlas $\mathcal{A}$. Given a coordinate chart $(x,\Omega)$ in $\A$, we let $\Gamma_x$ denote connection components represented in $x$-coordinates, and we let $\Gamma$ denote the collection of all representations $\Gamma_x$ of $\A$, i.e., $\Gamma \equiv \big\{ \Gamma_x \mid (x,\Omega) \in \A \big\}$. For simplicity always refer to $\Gamma$ as an (affine) connection; note, (though never used in this paper), in our notation $\nabla_x \equiv \partial_x + \Gamma_x$ defines a covariant derivative on the tangent bundle of $\M$ represented in $x$-coordinates.     An affine connection $\Gamma$, defined on $(\M,\A)$, is said to have regularity $W^{s,p}$, if the components of $\Gamma$ have regularity $W^{s,p}$ in every coordinate patch $(x,\Omega)$ of $\mathcal{A}$, in which case we write $\Gamma \in W^{s,p}_{\mathcal{A}}$, and for components of $\Gamma$ in a fixed coordinate chart $(x,\Omega)$ we write $\Gamma_x \in W^{s,p}(\Omega_x)$ with $\Omega_x \equiv x(\Omega) \subset \R^n$.  Similarly, we write $\Riem(\Gamma)\in W^{s,p}_{\mathcal{A}}$ to express that the components of the Riemann curvature tensor have regularity $W^{s,p}$ in each coordinate system $(x,\Omega)$ of the atlas $\mathcal{A}$, and $\Riem(\Gamma_x) \in W^{s,p}(\Omega_x)$ denotes the regularity of curvature components in a fixed coordinate chart $(x,\Omega)$. We let $\A_s$ denote an atlas which preserves $W^{s,p}$ connection regularity, (so $\A_s$ can be any atlas with transition maps of regularity $W^{2,p}$ or higher).\footnote{It is well-known that Sobolev norms are typically non-invariant in non-Riemannian geometry, i.e., their values vary with the coordinate system or depend on additional structures, like a Riemannian background metric. In the context of this paper such a background metric can only be maintained in $W^{1,p}$, a regularity too low to be of any use. Sobolev spaces, on the other hand, have invariant meaning in the sense that if a tensor or connection is in $W^{s,p}$ in one coordinate chart, then it will also be in $W^{s,p}$ in any other coordinate chart within any atlas of sufficiently regular transition maps, $W^{s+1,p}$ for tensors, and $W^{s+2,p}$ for connections.   }        Our first lemma establishes that the regularity of a connection controls the regularity of an atlas as follows:

\begin{Lemma} \label{Lemma_atlas-control}
If $\Gamma\in W^{s,p}_{\mathcal{A}}$, then all transition maps of $\mathcal{A}$ have regularity $W^{s+2,p}$. 
\end{Lemma}

Given an atlas $\A$ with transition maps of regularity $W^{s,p}$, we define $\A^{\text{max}(s)}$ to be the extension of $\A$ to the maximal $W^{s,p}$ atlas, that is, $\A^{\text{max}(s)}$ is the maximal collection of all coordinate charts which contain the original atlas $\A$ and have $W^{s,p}$ transition maps on their overlaps. An $W^{s',p}$ atlas $\A$ can always be extended to a unique maximal $W^{s,p}$ atlas $\A^{\text{max}(s)}$, whenever $s'\geq s$, cf. \cite{Hirsch}.  By the connection transformation law, connection regularity $W^{s,p}$ is preserved within any $W^{s+2,p}$ atlas $\A$, is preserved under restriction to higher regularity subatlases, and is preserved under any $W^{s+2,p}$ extension of $\A$, including to the unique maximal $W^{s+2,p}$ atlas $\A^{\text{max}(s+2)}$.  

Our theory of the RT-equations shows that regularizing a $W^{s,p}$ connection requires a coordinate transformation of precisely the regularity $W^{s+2,p}$, and no more regular \cite{ReintjesTemple_ell3, ReintjesTemple_ell4}. From this we deduce that to regularize a connection $\Gamma \in W^{s,p}_\A$, requires extending the given atlas $\A$ by $W^{s+2,p}$ coordinate transformations in order to find a subatlas in which the connection exhibits a higher regularity. From this we conclude that a regularizing atlas must lie within the $W^{s+2,p}$ extension of $\A$.   For example, if a connection is represented at regularity $W^{s,p}$ in a given $C^\infty$ atlas, then to regularise $\Gamma$ one would need to go down from the $C^\infty$ atlas to the maximal $W^{s+2,p}$ extension of that $C^\infty$ atlas, to find a subatlas in which the connection is more regular than $W^{s,p}$. More generally, consider now the case of connection $\Gamma \in L^p$, the lowest connection regularity we address. In this case, to regularize $\Gamma$ requires finding subatlases within the maximal $W^{2,p}$ extension of our given atlas. Based on this, and the observation that the maximal $W^{2,p}$ extension of an atlas contains all $W^{s,p}$ extensions for any $s\geq 2$, we define the essential regularity of a connection in terms of the unique maximal $W^{2,p}$  extension $\A^{\text{max}(2)}$ of a given atlas $\mathcal{A}$, which we denote as $\Amax \equiv \A^{\text{max}(2)}$.

\begin{Def}\label{Def1}
We say a connection $\Gamma$ defined on ($\mathcal{M},\mathcal{A}$) has global essential regularity $m\equiv ess_{\mathcal{M}}(\Gamma)\geq 0$, $m\in \mathbb{N}$, if there exists a subatlas $\mathcal{A}_m$ of the maximal $W^{2,p}$ atlas $\Amax$ of $\A$ such that $\Gamma\in W^{m,p}_{\mathcal{A}_m},$  and there does not exist a subatlas $\mathcal{A}_s$ of $\Amax$ in which $\Gamma\in W^{s,p}_{\mathcal{A}_s}$ with $s\in \mathbb{N}$ and  $s>m$.        
\end{Def} 

\noindent Note that by Lemma \ref{Lemma_atlas-control}, $\Gamma\in W^{m,p}_{\mathcal{A}_m}$ with $m=ess_{\mathcal{M}}(\Gamma)$ implies that all transition maps of the atlas $\mathcal{A}_m$ have regularity $W^{m+2,p}$. 

Our necessary and sufficient condition for essential regularity asserts that $\Gamma\in W^{s,p}_{\mathcal{A}}$ exhibits its essential regularity $s=m \equiv ess_{\mathcal{M}}(\Gamma)\in \mathbb{N}_0$ if and only if the regularity of $\Riem(\Gamma)$ is precisely one derivative below the regularity of $\Gamma$ in the defining atlas $\mathcal{A}$;  namely, $\Riem(\Gamma)\in W^{s-1,p}_{\mathcal{A}}$, but $\Riem(\Gamma)\notin W^{s,p}_{\mathcal{A}}$. This is recorded in the following more general theorem, which establishes boundedness of the curvature (in suitable Sobolev spaces) as a definitive necessary and sufficient condition which determines when a (singular) connection can be regularized.  

\begin{Thm}\label{Thm1}  
Assume $\Gamma\in W^{s,p}_{\mathcal{A}_s}$ is given on an $n$-dimensional manifold ($\mathcal{M},\mathcal{A}_s$), $n\geq 2$, for $n<p<\infty$, $s\geq0$, (but $p>4$ in case $n\leq 3$ and $s=0$). Then:      
\begin{itemize}[leftmargin=.3in]  \setlength\itemsep{.5em}   
\item[{\bf (i)}] $\Gamma\in W^{s+1,p}_{\mathcal{A}_{s+1}}$ in some subatlas $\mathcal{A}_{s+1}$ of the maximal $W^{2,p}$ atlas $\Amax_s$ of $\A_s$ if and only if $\Riem(\Gamma)\in W^{s,p}_{\mathcal{A}_s}$.\footnote{\label{footnote_atlas} Note that $\mathcal{A}_{s+1}$ actually lies within the smaller atlas $\A_s^{\text{max}(s+2)}$, because the transition maps between $\mathcal{A}_s$ and $\mathcal{A}_{s+1}$ preserve connection regularity $W^{s,p}$, and by Lemma \ref{Lemma_atlas-control}, have regularity $W^{s+2,p}$. The subatlas $\mathcal{A}_{s+1}$ constructed via the RT-equations always lies in $\A_s^{\text{max}(s+2)} \subset \Amax_s$, (cf. Section \ref{Sec_global}). }     
\item[{\bf (ii)}]  If $ess_{\mathcal{M}}(\Gamma)<\infty$, then subsequent use of the RT-equations provides an algorithm for constructing an atlas $\A_m \subset \Amax$ in which $\Gamma$ exhibits its essential regularity $m=ess_{\mathcal{M}}(\Gamma)$. 
\end{itemize}
\end{Thm}  

\noindent If $ess_{\mathcal{M}}(\Gamma)=\infty$, we have the following result:

\begin{Corollary}  \label{Cor_RT->ess_reg}
Assume $\Gamma \in W^{s,p}_{\A_s}$ as in Theorem \ref{Thm1}.  If $ess_{\mathcal{M}}(\Gamma)=\infty$, then for each integer $0\leq s' <\infty$ subsequent use of the RT-equations yields an atlas $\A_{s'} \subset \Amax_s$ in which $\Gamma$ exhibits regularity $W^{s',p}_{\A_{s'}}$.               
\end{Corollary}

As a natural extension of the above results, we obtain the following theorem, which is a refinement of our necessary and sufficient condition and establishes the hierarchy of apparent singularities introduced in Section \ref{Sec_intro}.

\begin{Thm} \label{Cor1}  {\rm (Hierarchy of apparent singularities)}\ \ \ 
Assume $\Gamma\in W^{s,p}_{\mathcal{A}}$ is given on an $n$-dimensional manifold ($\mathcal{M},\mathcal{A}$), $n\geq 2$, for $n<p<\infty$, $s\geq0$,   (but $p>4$ in case $n\leq 3$ and $s=0$), and assume $\A=\A^{\text{max}(s+2)}$ is maximal. Then: 
{\small
\begin{itemize}[leftmargin=.23in]  \setlength\itemsep{.5em}      
\item[{\bf (1)}] $ess_\M(\Gamma)=s$ if and only if $\Riem(\Gamma)\in W^{s-1,p}_{\mathcal{A}}$ and $\Riem(\Gamma) \not\in   W^{s,p}_{\mathcal{A}}$;
\item[{\bf (2)}] $ess_\M(\Gamma)=s+1$ if and only if $\Riem(\Gamma)\in W^{s,p}_{\mathcal{A}}$ and $\Riem(\Gamma) \not\in W^{s+1,p}_{\mathcal{A}}$;
\item[{\bf (3)}]  $ess_\M(\Gamma)\geq s+2$ if and only if $\Riem(\Gamma)\in W^{s+1,p}_{\mathcal{A}}$.  
\end{itemize}
}
\end{Thm} 

\noindent If $\ess_\M(\Gamma)=\infty$ we have the following immediate corollary of Theorem \ref{Cor1}.

\begin{Corollary}  \label{Cor1_infty}
We have $ess_{P}(\Gamma)=\infty$ if and only if for any chart $(x,\Omega) \in \A$, $\Gamma_x \in W^{s',p}(\Omega_x)$ implies $\Riem(\Gamma_x) \in W^{s'+1,p}(\Omega_x)$. 
\end{Corollary}

The proof of Theorem \ref{Cor1} is based on an asymmetry between the connection transformation law and the tensor transformation law for the curvature, noting that only the former involves derivatives of the Jacobian.  To the authors' knowledge, Theorem \ref{Cor1} gives the first definitive theory for determining when a singularity, interpreted as a connection represented below its essential regularity, is removable or not. As far as we know, before this, there were no definitive conditions for determining whether or not connections with $L^p$ components could be regularized by coordinate transformation, nor was the essential regularity of a connection understood to be a geometric property of connections.

\subsection{Discussion of Theorem \ref{Thm1}}

Part (i) of Theorem \ref{Thm1} asserts that a connection defined by an atlas $\mathcal{A}_s$ globally exhibits one additional derivative of regularity in a transformed subatlas $\mathcal{A}_{s+1}$ which lies in $\Amax_s$, (the maximal $W^{2,p}$ extension of $\A_s$), if and only if $\Riem(\Gamma)$ has the same regularity as $\Gamma$ in $\mathcal{A}_s$, i.e., $\Riem(\Gamma) \in W^{s,p}_{\A_s}$. Conversely, Theorem \ref{Thm1} asserts that if $\Riem(\Gamma)$ is one derivative {\it less} regular than $\Gamma$ in $W^{s,p}$ in atlas $\mathcal{A}_s$, then no subatlas exists within the entire maximal $W^{2,p}$ extension $\Amax_s$ of $\A_s$ in which the connection is one derivative more regular than $W^{s,p}$.

Part (ii) of Theorem \ref{Thm1} determines an algorithm for lifting any connection to its essential regularity by use of the RT-equations, provided that $ess_{\mathcal{M}}(\Gamma)<\infty$. To explain this, start with a connection $\Gamma$ with regularity $W^{s_0,p}$ below its essential regularity, defined in a given initial atlas $\mathcal{A}_{s_0}$, so $\Gamma\in W^{s_0,p}_{\mathcal{A}_{s_0}}$, $s_0\geq0,$ $p>n.$  By Theorem \ref{Thm1} (i),  $\Riem(\Gamma)\in W^{s_0,p}_{\mathcal{A}_{s_0}}$ implies the existence of atlas $\mathcal{A}_{s_0+1}$ in the maximal $W^{2,p}$ extension $\Amax_{s_0}$ of $\mathcal{A}_{s_0}$,\footnote{As discussed in Footnote \ref{footnote_atlas}, we actually have $\A_{s_0+1} \subset \A^{\text{max}(s_0+2,p)}_{s_0}\subset \Amax_{s_0}$, but to keep the discussion simple we consider here $\Amax$.} constructed by solving the RT-equations,  such that $\Gamma\in W^{s_0+1,p}_{\mathcal{A}_{s_0+1}}$. Letting $\Gamma_s$ and $\Riem(\Gamma_s)$ denote the components of $\Gamma$ and $\Riem(\Gamma)$ in atlas $\mathcal{A}_s$, respectively, and  applying Theorem \ref{Thm1} to $\Gamma_{s_0+1},$ we see that $\Gamma_{s_0+1}$ can be further regularized by coordinate transformation if and only if the components $\Riem(\Gamma_{s_0+1})$ have the regularity of components $\Gamma_{s_0+1}$, i.e., if and only if $\Riem(\Gamma_{s_0 +1})\in W^{s_0+1,p}_{\mathcal{A}_{s_0+1}}$.   If so, Theorem \ref{Thm1} (i) asserts that a new subatlas $\mathcal{A}_{s_0+2}$ of $\Amax_{s_0}$ can be constructed from $\mathcal{A}_{s_0+1}$ by solving the RT-equations, such that $\Gamma\in W^{s_0+2,p}_{\mathcal{A}_{s_0+2}}$.  It follows, then, that this procedure continues until one reaches the maximal value $s=m\equiv ess_{\mathcal{M}}(\Gamma)\geq s_0$, which by Theorem \ref{Thm1} must be the first level of regularity $W^{m,p}$, $m\geq s_0$, at which $\Riem(\Gamma_m)$ is one full derivative {\it less} regular than $\Gamma_m$ in atlas $\mathcal{A}_m$.      

Theorem \ref{Cor1} asserts that this value of $m$ will always be the essential regularity of $\Gamma$ as defined in Definition \ref{Def1}; and the composition of coordinate transformations determined by the RT-equations at each step $s=s_0,...,m$, provides coordinate transformations which lift the connection $\Gamma$ defined in the original atlas $\mathcal{A}_{s_0}$, directly to the atlas $\mathcal{A}_m$ in which $\Gamma$ achieves its essential regularity.   In this case, for each $s=s_0,...,m,$ the atlas $\mathcal{A}_{s}$ is a subatlas of the maximal $W^{2,p}$ extension $\Amax_{s_0}$ of the original atlas $\mathcal{A}_{s_0}$, and the regularity of atlas $\mathcal{A}_{s}$ will be $W^{s+2,p}$, the lowest regularity required to preserve the $W^{s,p}$ regularity of $\Gamma_s$ under coordinate transformations within the atlas $\mathcal{A}_{s}$.       In the case when $ess_{\mathcal{M}}(\Gamma)=m=\infty$, this procedure will regularize the connection components to $W^{s,p}$ for every integer $s< \infty$, but, since the existence theory for the RT-equations requires restrictions to smaller domains, not necessarily all the way to $s=\infty$.

\subsection{Hierarchy of $C^\infty$ differentiable structures} 

If we were to carry out the above theory of essential regularity for connections in the simpler classical spaces $C^k$, then every connection starting in $C^1$ would determine an essential (highest possible) regularity $C^m$ achieveable by $C^k$ coordinate transformations, and this would determine a sequence of $C^{s+2}$ atlases $\A_s$ in which the connection exhibits the regularity $C^s$, $1\leq s\leq m,$ and without loss of generality we can assume these are maximal $C^s$ extensions.   In this case Theorem 2.9 in \cite{Hirsch} applies, asserting that every maximal $C^s$ atlas contains a $C^\infty$ subatlas $\A^{\infty}_s$, and in this $C^\infty$ atlas the connection would continue to exhibit regularity $C^s$.   By the proof of Lemma \ref{J_derivatives} below, $\mathcal{A}_{s}^\infty$ and $\mathcal{A}_{s'}^\infty$ could be shown to be related by coordinate transformations which have the regularity $C^{s+2}$, $s\leq s'$, not $C^\infty$. By this we have a sequence of $C^{\infty}$ atlases which are not $C^{\infty}$ equivalent in which the connection exhibits its different possible regularities. Thus every connection determines a hierarchy of distinct $C^\infty$ differentiable structures (atlases), distinguished by the regularity of the connection.  We note here that this does not contradict \cite[Thm. 2.9]{Hirsch}, which goes on to assert that these $C^\infty$ atlases are all $C^\infty$ diffeomeorphic, because, in contrast to coordinate mappings, the $C^\infty$ diffeomorphisms used in \cite{Hirsch} are constructed by mollification, and hence the image of the diffeomorphisms does not represent the connection at the same point as the pre-image. Note that the existence of such $C^{\infty}$ atlases implies that one loses no generality by restricting to the set of connections which exhibit essential regularity in a $C^{\infty}$ atlas.     

The authors find it interesting that naively one might think that the regularity of a manifold is given by the regularity of its atlas, but this cannot be a geometric property of manifolds alone, because atlases can be extended to a unique maximal $C^1$ atlas, and all subsequent restrictions to $C^\infty$ subatlases are $C^\infty$ equivalent in the sense of \cite[Thm. 2.9]{Hirsch}. Thus a manifold alone does not have enough structure to determine a geometric level of regularity. Our results here establish that a manifold together with a connection does have an inherent geometric level of regularity, the essential regularity of the connection.

\section{The RT-equations} \label{Sec_RT}

In this section we establish a refined existence theory for the RT-equations which is required for the results in this paper. We begin by recording the authors' prior results on the {\it Regularity Transformation (RT-) equations}, developed in a series of papers \cite{ReintjesTemple_ell1, ReintjesTemple_ell2, ReintjesTemple_ell3, ReintjesTemple_ell4, ReintjesTemple_ell5, ReintjesTemple_ell6, ReintjesTemple_geo}, as required for this paper. We proved in \cite{ReintjesTemple_ell2, ReintjesTemple_ell4} that solutions of the RT-equations furnish local coordinate transformations which regularize connections to one derivative of regularity above their Riemann curvature, as measured in a starting coordinate system. In prior papers we were content with the one step regularization of connections to one derivative above the initial curvature regularity, and considered the generic case when the curvature regularity stays fixed under the regularizing transformation. From the point of view of the present paper, we did not take account of the fact that the curvature is regularized along with the connection when connection regularity is more than one derivative below essential regularity, as asserted by Theorem \ref{Thm1} above. In our prior papers we did not fully understand that there can be an {\it automatic} regularization of the curvature and that subsequent use of the RT-equations would take one to an essential, i.e., best possible, {\it geometric} regularity of a connection.\footnote{In this paper we thus distinguish ``essential regularity'' from ``optimal regularity'', the latter being standard terminology  (cf. \cite{KazdanDeTurck, Anderson}) we used in our prior papers. That is, we said a connection has been lifted to ``optimal regularity'' by coordinate transformation if it exhibits one derivative of regularity above its Riemann curvature in some starting coordinate system--a coordinate dependent terminology which is imprecise in light of our new results here.}  

To state our theorems in \cite{ReintjesTemple_ell2,ReintjesTemple_ell4}, which addressed only the local regularization of connections, consider a fixed chart $(x,\Omega)$ on an $n$-dimensional manifold $\mathcal{M}$, where $\Omega_x \equiv x(\Omega) \subset \R^n$ (the image of $\Omega$ under the coordinate map)  is open and bounded with smooth boundary. Let $\Gamma_x$ denote the collection of components $\Gamma^k_{ij}(x)$ of an affine connection $\Gamma$ in $x$-coordinates.  We view $\Gamma_x$ as a matrix valued $1$-form in $x$-coordinates, $(\Gamma_x)^\mu_{\nu} \equiv (\Gamma_x)^\mu_{\nu j} dx^j$, and we use the Einstein summation convention of summing over repeated indices from $1$ to $n$. Let $d\Gamma_x$ denote its exterior derivative, $d(\Gamma_x)^\mu_\nu \equiv \partial_i (\Gamma_x)^\mu_{\nu j} dx^i \wedge dx^j$, where $\mu, \nu = 1,...,n$ denote indices of the matrix. Writing the Riemann curvature tensor as a matrix valued $2$-form, ${\rm Riem}(\Gamma_x) = d\Gamma_x +\Gamma_x \wedge \Gamma_x$, it follows that the assumption $\Gamma_x \in L^{p}(\Omega_x)$ and ${\rm Riem}(\Gamma_x) \in L^{p/2}(\Omega_x)$ is equivalent to the assumption $\Gamma_x \in L^{p}(\Omega_x)$ and $d\Gamma_x \in L^{p/2}(\Omega_x)$, and we assume the latter. The main idea for establishing optimal regularity in \cite{ReintjesTemple_ell4} was to derive from the connection transformation law a non-invariant system of {\it elliptic} PDE's for the regularizing Jacobian $J$, an idea motivated by the Riemann-flat condition in \cite{ReintjesTemple_geo}. This idea led to our formulation of the {\it RT-equations} in \cite{ReintjesTemple_ell1}.

\subsection{Derivation of the RT-equations}
We now give the main steps in the derivation of the RT-equations, and explain how they furnish optimal regularity. For this, assume there exists a coordinate transformation with Jacobian $J$ mapping $\Gamma_x$ to $\Gamma_y$ (the connection of optimal regularity), and write the connection transformation law as
\beq \label{opt_eqn1}
\Gammati = \Gamma - J^{-1} dJ,
\eeq
where $\Gamma \equiv \Gamma_x$ and $\Gammati^k_{ij} = (J^{-1})^k_\gamma J^\alpha_i J^\beta_j (\Gamma_y)^\gamma_{\alpha\beta}$ is the connection $\Gamma_y$ transformed as a tensor to $x$-coordinates. Differentiating \eqref{opt_eqn1} by the exterior derivative $d$ and by the co-derivative $\delta$, implies the following two equations
\begin{eqnarray} 
\Delta \Gammati &=& \delta d\Gamma - \delta\big(dJ^{-1} \wedge dJ\big) + d\delta \Gammati ,     \label{opt_eqn2} \\
\Delta J &=& \delta(J\Gamma) - \langle dJ ; \Gammati \rangle - J \delta\Gammati ,  \label{opt_eqn3}
\end{eqnarray}
where $\Delta \equiv \delta d + d \delta = \partial_{x^0}^2 + ... + \partial_{x^n}^2$ is the Euclidean Laplacian, $\langle \cdot\; ; \cdot \rangle$ is a matrix-valued inner product and $\wedge$ the wedge product on matrix valued differential forms, (see \cite[Ch.3]{ReintjesTemple_ell1} or \cite[Ch.5]{ReintjesTemple_ell4} for detailed definitions). At this stage, equations \eqref{opt_eqn2} - \eqref{opt_eqn3} neither appear solvable, nor need a solution $J$ be a true Jacobian that is integrable to coordinates, i.e. satisfying ${\rm Curl}(J) =0$. To complete the equations, view $A\equiv \delta\Gammati$ as a free matrix valued function---this choice was motivated by the Riemann-flat condition for optimal regularity in \cite{ReintjesTemple_geo}, because the latter only involves $d\Gammati$, but not $\delta\Gammati$.     Substituting $A\equiv \delta\Gammati$ in \eqref{opt_eqn2} - \eqref{opt_eqn3}, and viewing $A$ as a new unknown matrix valued function, we next impose on equation \eqref{opt_eqn3} the condition ${\rm Curl}(J) =0$ for integrability.  For this, we introduce the vectorization $\vec{J}^\mu = J^\mu_\nu dx^\nu$ of $J$, so that ${\rm Curl}(J)\equiv d\vec{J}$, and impose equivalently $d\vec{J} =0$ as the integrability condition.   By a fortuitous cancellation the regularities in different terms of the equation become consistent, and the computations in \cite{ReintjesTemple_ell1} eventually lead to the {\it RT-equations}:
\begin{align} 
\Delta \Gammati &= \delta d\Gamma - \delta \big(d J^{-1} \wedge dJ \big) + d(J^{-1} A ), \label{PDE1} \\
\Delta J &= \delta ( J \Gamma ) - \langle d J ; \tilde{\Gamma}\rangle - A , \label{PDE2} \\
d \vec{A} &= \overrightarrow{\text{div}} \big(dJ \wedge \Gamma\big) + \overrightarrow{\text{div}} \big( J\, d\Gamma\big) - d\big(\overrightarrow{\langle d J ; \tilde{\Gamma}\rangle }\big),   \label{PDE3}\\
\delta \vec{A} &= v.  \label{PDE4}
\end{align}

Equation \eqref{PDE3} on the auxiliary field $A$ results from imposing $d\vec{J}=0$ on \eqref{PDE2}, and one can prove integrablity of $J$ follows from the coupled equations \eqref{PDE2} and \eqref{PDE3}. The unknowns $(\Gammati,J,A)$ in the RT-equations, together with the given non-optimal connection components $\Gamma$, are viewed as matrix valued differential forms. Arrows denote ``vectorization'', mapping matrix valued $0$-forms to vector valued $1$-forms, (e.g. $\vec{A}^\mu = A^\mu_i dx^i$) and $\overrightarrow{\text{div}}$ is a divergence operation which maps matrix valued $k$-forms to vector valued $k$-forms. The vector $v$ in \eqref{PDE4} is free to impose, representing a ``gauge''-type freedom in the equations, reflecting the fact that smooth transformations preserve optimal connection regularity. The operations on the right hand side are formulated in terms of the Cartan Algebra of matrix valued differential forms based on the Euclidean metric in $x$-coordinates, and these objects depend on the starting coordinate system and are not invariant under change of coordinates, (see \cite{ReintjesTemple_ell1} for detailed definitions and proofs).  As we show in this paper, a regularity below essential regularity is not a geometric property of a connection, so it makes sense that non-invariant equations are required to regularize them. The RT-equations \eqref{PDE1} - \eqref{PDE4} are a non-invariant solvable system of PDE's which is elliptic regardless of metric signature.

\subsection{How the RT-equations yield optimal regularity}

We now explain how solutions of the RT-equations furnish the local coordinate transformations to optimal regularity, leading to our main Theorems 2.5 and 3.1 in \cite{ReintjesTemple_ell2} and \cite{ReintjesTemple_ell4}, respectively. Because our earlier iteration scheme for solving the RT-equations at higher regularity in \cite{ReintjesTemple_ell2,ReintjesTemple_ell3} did not close at the low regularity of $L^p$ connections, due to the non-linear term $d J^{-1} \wedge dJ$ in \eqref{PDE1}, we eventually discovered an internal ``gauge''-type transformation for solutions of the RT-equations \eqref{PDE1} - \eqref{PDE4}, and by this we succeeded in separating off the troublesome equation \eqref{PDE1} from the remaining equations, c.f. \cite{ReintjesTemple_ell4}. The resulting system is linear in the unknowns $(J,B)$, and we refer to this system as the {\it reduced} RT-equations which take the form \cite{ReintjesTemple_ell4}: 
\begin{eqnarray} 
\Delta J &=& \delta ( J \mm \Gamma ) - B , \label{RT_withB_2} \\
d \vec{B} &=& \overrightarrow{\text{div}} \big(dJ \wedge \Gamma\big) + \overrightarrow{\text{div}} \big( J\, d\Gamma\big) ,   \label{RT_withB_3} \\
\delta \vec{B} &=& v'.  \label{RT_withB_4}
\end{eqnarray}
Our iteration scheme, which is based on solving the linear Poisson equation at each stage, applies to the reduced RT-equations \eqref{RT_withB_2} - \eqref{RT_withB_4} at the low regularity of $L^p$ connections with $d\Gamma \in L^{p/2}$, and locally establishes existence of solutions $(J,B)$, (i.e, in neighborhoods $\Omega'$ of points), such that $J$ is point-wise an invertible matrix, cf. \cite[Thm 6.4]{ReintjesTemple_ell4}. 

That any solution $J$ is a Jacobian integrable to coordinates is a built-in property of \eqref{RT_withB_2} - \eqref{RT_withB_4}, provided that the integrability condition $d\vec{J}\equiv {\rm Curl}(J)=0$ holds on the boundary $\partial\Omega'$, c.f. \cite[Thm 6.4]{ReintjesTemple_ell4}. That is, combining \eqref{RT_withB_2} with \eqref{RT_withB_3}, a computation shows that $\omega\equiv d\vec{J}$ is a solution of Laplace's equation $\Delta \omega=0$, which together with our boundary data implies that $\omega=0$ throughout $\Omega'$. This implies that $J$ is integrable to coordinates. 

Given now a solution $(J,B)$ of the reduced RT-equation \eqref{RT_withB_2} - \eqref{RT_withB_4} with $J$ an integrable and invertable Jacobian, one recovers a solution $(J,\Gammati,A)$ of the full RT-equations \eqref{PDE1} - \eqref{PDE4} by introducing\footnote{The second and third equation in \eqref{Gammati'} define the ``gauge'' transformations of the RT-equations, while the first equation defines a projection onto the space of solution of the Riemann-flat condition.}
\beq  \label{Gammati'} 
\Gammati \equiv \Gamma - J^{-1} dJ, 
\hspace{.5cm} 
A \equiv B - \langle d J ; \Gammati \rangle ,  
\hspace{.5cm} \text{and} \hspace{.5cm}
v \equiv v' - \delta \overrightarrow{\langle d J ; \Gammati \rangle},
\eeq
as can be verified by direct computation using \eqref{RT_withB_2} to eliminate uncontrolled terms involving $\delta\Gamma$. From interior elliptic estimates, applied to the first RT-equations \eqref{PDE1}, one can then prove that $\Gammati$ is in $W^{1,p}$ on any open set $\Omega'_c$ compactly contained in $\Omega'$, a gain of one derivative over $\Gamma$. Defining 
\beq \label{Gamma_y_reverse}
(\Gamma_y)^\gamma_{\alpha\beta} \equiv J_k^\gamma (J^{-1})^i_\alpha  (J^{-1})^j_\beta   \; \Gammati^k_{ij},
\eeq
substitution into the first equation in \eqref{Gammati'} yields the connection transformation law \eqref{opt_eqn1}, which implies that $\Gamma_y$ is the transformed connection of optimal regularity, $\Gamma_y \in W^{1,p/2}(\Omega'_c)$. 

Based on these ideas, we proved in \cite{ReintjesTemple_ell4} our optimal regularity result, for $L^p$ connections with Riemann curvature in $L^{p/2}$, (cf. Theorem 3.1. in \cite{ReintjesTemple_ell4}). Before we state this result precisely, we give a restatement our prior optimal regularity result Theorem 2.5 in \cite{ReintjesTemple_ell3} adapted to our setting here, which addresses the easier case of higher non-optimal connection regularity $W^{s,p}$ for $s\geq 1$ and $p>n$. This is an easier problem, because the non-linear gradient product in \eqref{PDE1} can be controlled by Morrey's inequality in our iteration scheme. Moreover, in this case $\Gamma$ can be taken to be in the same Sobolev space $W^{s,p}$ as $\Riem(\Gamma)$, because the wedge product $\Gamma \wedge \Gamma$ has the same regularity as $\Gamma$, again by Morrey's inequality.    Recall, our notation is that $\Gamma_x$ denotes the components of connection $\Gamma$ in a coordinate system $x$ defined on $\Omega_x\equiv x(\Omega)\subset{\mathbb R}^n$ for $\Omega\subset\mathcal{M}$ open, $\Gamma_y$ denotes the components of the regularized connection in $y$ coordinates on $\Omega_y\equiv y(\Omega)$, and regularity is measured component-wise in coordinate dependent norms. 

\begin{Thm}  \label{Thm_Smoothing_high}
Assume the components of $\Gamma_x, \Riem(\Gamma_x) \in W^{s,p}(\Omega_x)$ for $s\geq 1$,  $p\in (n,\infty)$, $n\geq2$, in $x$-coordinates, such that 
\beq \label{bound_incoming_high}
\|(\Gamma,\Riem(\Gamma))\|_{W^{s,p}(\Omega_x)} 
\equiv \|\Gamma_x\|_{W^{s,p}(\Omega_x)} + \|\Riem(\Gamma_x)\|_{W^{s,p}(\Omega_x)} 
\leq M,
\eeq  
for some constant $M\geq 0$. Then for each $P\in\Omega$ there exists a neighborhood $\Omega'\subset \Omega$ of $P$ (depending only on $M$ and $\Omega_x, s, n, p$) and a coordinate transformation $x \to y$ with Jacobian $J\equiv \frac{\partial y}{\partial x} \in W^{s+1,p}(\Omega'_x)$, such that $\Gamma_y \in W^{s+1,p}(\Omega'_y)$ and 
\beq \label{curvature_estimate_Main}
\|\Gamma_y\|_{W^{s+1,p}(\Omega'_y)}  \leq C(M) ,
\eeq
and the Jacobian $J$ and its inverse $J^{-1}$ (expressed in $x$-coordinates) satisfy
\beq \label{curvature_estimate_J}
\|J\|_{W^{s+1,p}(\Omega'_x)}  + \|J^{-1}\|_{W^{s+1,p}(\Omega'_x)}  \leq C(M),
\eeq 
where $C(M)>0$ is some constant depending only on $M$ together with $s, n, p, \Omega_x$.\footnote{In \cite{ReintjesTemple_ell4} we state refined version of estimates \eqref{bound_incoming_high} and \eqref{curvature_estimate_Main} in terms of $d\Gamma$ in place of $\Riem(\Gamma)$. Expressing the Jacobians in $y$-coordinates yields the same bound, but with a potentially different constant $C(M)$.}
\end{Thm}

Estimate \eqref{curvature_estimate_Main} is the uniform bound from which we derive Uhlenbeck compactness in \cite{ReintjesTemple_ell3}. We now restate Theorem 3.1 in \cite{ReintjesTemple_ell4} which addresses the low regularity case of $L^p$ connections. Its proof is worked out in detail at the level of weak solutions in \cite{ReintjesTemple_ell4}. 

\begin{Thm} \label{Thm_Smoothing_low}  
Assume $\Gamma_x \in L^{p}(\Omega_x)$ and $\Riem(\Gamma_x) \in L^{p/2}(\Omega_x)$ in $x$-coordinates, for some $p>{\rm max}\{4 ,n\}$, $n\geq2$. Then for any point $P\in \Omega$ there exists a neighborhood $\Omega' \subset \Omega$ of $P$ (depending on $\Omega_x, n, p$ and $\Gamma$) and a coordinate transformation $x \to y$ with Jacobian $J=\frac{\partial y}{\partial x}\, \in W^{1,p}(\Omega'_x)$, such that $\Gamma_y \in W^{1,p/2}(\Omega'_y)$. 
Moreover, if in addition $\Gamma_x \in L^\infty(\Omega_x)$ and $\Riem(\Gamma_x) \in L^{p}(\Omega_x)$ such that\footnote{In this case we only need to require $p>n\geq 2$, but not $p>4$.}    \beq \label{bound_incoming_ass}
\|(\Gamma,\Riem(\Gamma) ) \|_{L^{\infty, p}(\Omega_x)} \equiv 
\|\Gamma_x \|_{L^{\infty}(\Omega_x)} + \|\Riem(\Gamma_x) \|_{L^{p}(\Omega_x)} \; \leq \; M
\eeq    
for some constant $M>0$, then $\Omega'$ depends only on $M$ and $\Omega_x, n, p$, but not on $\Gamma$ near $P$, and $\Gamma_y$ satisfies   
\beq \label{bound_optimal_reg1} 
\|\Gamma_y \|_{W^{1,{p}}(\Omega'_y)}  \leq C(M), 
\eeq
and the Jacobian $J$ and its inverse $J^{-1}$ (expressed in $x$-coordinates) satisfy
\beq \label{bound_optimal_reg2} 
\|J\|_{W^{1,{2p} }(\Omega'_x)}  + \|J^{-1}\|_{W^{1,{2p} }(\Omega'_x)} \leq C(M),
\eeq
for some constant $C(M) > 0$ depending on $M$ and $n, p, \Omega_x$.   
\end{Thm}

For example, if $\Omega'$ is taken to be a sequence of balls of radius $r(M)$ centered at $P$, and $C(M)$ is taken large enough so that $r(M)<C(M)^{-1}$, then Theorem \ref{Thm_Smoothing_low} asserts that there is a single function $C(M)$ which tells how small a neighborhood restriction $\Omega'$ is required to bound connections uniformly in the higher norm $W^{1,p}$, given they meet the incoming bound \eqref{bound_incoming_ass} in the lower norm $L^\infty$. From this we conclude Uhlenbeck compactness in \cite{ReintjesTemple_ell4}. 

Theorem \ref{Thm_Smoothing_low} implies the connection regularity can be lifted from $\Gamma_x\in L^p$ to $\Gamma_y\in W^{1,p/2},$ which is one derivative more regular than the starting curvature $\Riem(\Gamma_x)\in L^{p/2}$.   For the purposes of our present paper, based on levels of regularity which we view as resulting from singular coordinate transformations of more regular connections, we need to conclude the existence of coordinate transformations which lift $\Gamma_x\in L^p$ to $\Gamma_x\in W^{1,p}$ under the stronger assumption $\Riem(\Gamma_x)\in L^p$, in place of $\Riem(\Gamma_x)\in L^{p/2}$, that is, we need to assume the same value of $p>n$ for the curvature and the connection. The following corollary, which is new to this paper, establishes this result by consecutive use of the regularization asserted in Theorem \ref{Thm_Smoothing_low} in combination with Sobolev embedding:

\begin{Corollary} \label{Thm_Smoothing_low-cor}  
Assume $\Gamma_x \in L^{p}(\Omega_x)$ and $\Riem(\Gamma_x) \in L^{p}(\Omega_x)$ in $x$-coordinates, for some $p>{\rm max}\{4,n\}$. Then for any point $P\in \Omega$ there exists a neighborhood $\Omega' \subset \Omega$ of $P$ (depending on $\Omega_x, n, p$ and $\Gamma$) and a coordinate transformation $x \to y$ with Jacobian $J, J^{-1} \in W^{1,p}(\Omega'_x)$,  such that $\Gamma_y \in W^{1,p}(\Omega'_y)$.   
\end{Corollary}

\Proof   
Assume $\Gamma_x\in L^{p}$ and $\Riem(\Gamma_x)\in L^p$, for $p>{\rm max}\{4,n\}$. This implies the weaker assumption $\Gamma_x\in L^{p}$ and $\Riem(\Gamma_x)\in L^{p/2}$ of Theorem \ref{Thm_Smoothing_low}. So Theorem \ref{Thm_Smoothing_low}  implies the existence of a coordinate transformation $x \to y'$ defined on $\Omega'\subset\Omega$ containing $P$, such that the Jacobian $J \in W^{1,p}$ transforms $\Gamma_x$ to a connection $\Gamma_y'\in W^{1,p/2}$.  Since $J \in W^{1,p}$ and $p>n$, Morrey's inequality implies that $J$ is H\"older continuous,\footnote{Morrey's inequality states that $\|f\|_{C^{0,\alpha}} \leq C \|f\|_{W^{1,p}}$ for $p>n$ and $\alpha=1-\frac{n}{p}$, for some constant $C>0$, \cite[Ch.5]{Evans}. We mainly use this to bound $L^\infty$ norms to control non-linear products.} and this regularity of $J$ preserves the regularity of the curvature under the tensor transformation law, 
\beq \label{Riemann_transfo_Sec3}
\Riem(\Gamma_x)^i_{jkl}
= (J^{-1})^i_\delta \; J^\alpha_j J^\beta_k J^\gamma_l \; \,\Riem(\Gamma_y)^\delta_{\alpha \beta \gamma}, 
\eeq
so $\Riem(\Gamma_{y'}) \in L^p$. If $p>2n$, this already suffices to establish Corollary \ref{Thm_Smoothing_low-cor}, because $\Gamma_{y'} \in W^{1,p/2} \subset L^\infty(\Omega')$ by Morrey's inequality, so that Theorem \ref{Thm_Smoothing_low} directly yields a coordinate transformation $y'\to y$ regularizing the connection to $W^{1,p}$ in some neighborhood of $P$. 

For the more problematic case $n < p \leq 2n$, note that by Sobolev embedding \cite{Evans}, $\Gamma_{y'} \in W^{1,p/2}$ implies $\Gamma_{y'} \in L^{\phi(p)}$ with estimate $\|\Gamma_{y'}\|_{L^{\phi(p)}} \leq C \|\Gamma_{y'}\|_{W^{1,p}}$ for some generic constant $C>0$, where $\phi(p)$ is the Sobolev conjugate of $p/2$, $\phi(p) \equiv \frac{pn}{2n-p}$. Now if $p \in [3/2n,2n]$, then $\phi(p)\geq 2p$, implying $\Gamma_{y'} \in L^{\phi(p)} \subset L^{2p}$. Thus, since $\Riem(\Gamma_{y'}) \in L^p$, we can directly apply Theorem \ref{Thm_Smoothing_low} to conclude the existence of a coordinate transformation $y'\to y$ regularizing the connection to $W^{1,p}$ in a neighborhood of $P$.  

Finally, if $p \in (n,3/2n)$, then $\phi(p) < 2p$, and regularization by the RT-equations only yields a connection in $W^{1,\phi(p)/2}$, still short of the requisite $\phi(p)\geq 2p$. However, since $\frac{d \phi}{dp}>0$ and $\frac{d^2 \phi}{dp^2}>0$ for all $p\in (n,2n)$, it follows that
$p < \phi(p) < \phi(\phi(p)) <...$, with increasing step size between each subsequent composition of $\phi$ with itself. This implies that after a finite number of successive application of the RT-equations in combination with Sobolev embedding (in the above fashion) yields a coordinate system $y''$ such that $\Gamma_{y''} \in L^{2p}$ and $\Riem(\Gamma_{y''}) \in L^p$. One can now apply again Theorem \ref{Thm_Smoothing_low} to conclude the existence of a coordinate transformation $y''\to y$ regularizing the connection to $W^{1,p}$ in a neighborhood of $P$. Taking $x\to y$ to be the composition of these maps, we conclude that there always exists a coordinate transformation $x\to y$ in a neighborhood of $P$ which lifts the regularity of the connection $\Gamma_x\in L^p$ to  $\Gamma_y\in W^{1,p}$, for every $p>n$, provided $\Riem(\Gamma_x)\in L^p$. This completes the proof of Corollary \ref{Thm_Smoothing_low-cor}. 
\QED

\section{The Link Between Connection and Jacobian Regularity}  \label{Sec_Jacobian}

The consistency of the notion of essential regularity hinges on a simple but surprisingly consequential lemma which identifies a rigid relationship between the regularity of transformed connection coefficients, and the regularity of the Jacobians which transform them.    To state this carefully, consider a connection represented by components $\Gamma_x$ and $\Gamma_y$ in two different coordinate systems $x$ and $y$, both defined on the same open set $\Omega$.  Let $J \equiv \frac{\partial y}{\partial x}$ denote the regular invertible Jacobian of the coordinate transformation $x\to y,$  so $J^{-1}$ exists and has the same regularity as $J$, (as follows by differentiating $J^{-1} J = I$).  Assume $\Gamma_y \in W^{s,p}$ and $\Gamma_x\in W^{r,p}$, (shorthand for $\Gamma_y \in W^{s,p}(\Omega_y)$ and $\Gamma_x\in W^{r,p}(\Omega_x)$), where  $r,s\geq0$, $p>n$, and without loss of generality assume $r\leq s.$  The following Lemma, which applies to any neighborhood $\Omega$ on which the mapping $y\to x$ is defined, implies that the regularity of $J$ is always {\it at least} one derivative above the regularity of $\Gamma_x,$ and is {\it exactly} one derivative above the regularity of $\Gamma_x$, whenever we have the precise regularity $\Gamma_x\in W^{r,p}$, $\Gamma_x\notin W^{r',p}$, any $r'>r$.     

\begin{Lemma} \label{Lemma_BigStart}
Let $J$ be the Jacobian of a coordinate transformation which transforms $\Gamma_y \in W^{s,p}$ to $\Gamma_x$, and let $0\leq r\leq s$ and $p>n$. Then the components of $\Gamma_x$ satisfy $\Gamma_x\in W^{r,p}$ if and only if the components of $J$ satisfy $J\in W^{r+1,p}$.
\end{Lemma}

\Proof 
This follows from the connection transformation law
\beq \label{connection_transfo}
(\Gamma_x)^\mu_{\rho\nu}
= (J^{-1})^\mu_\alpha \Big( J^\beta_\rho J^\gamma_\nu  \, (\Gamma_y)^\alpha_{\beta \gamma}   +  \tfrac{\partial }{\partial x^\rho}  J^\alpha_\nu \Big).
\eeq
For the reverse implication, if $J\in W^{r+1,p}$, then the right hand side of (\ref{connection_transfo}) is in $W^{r,p}$, implying $\Gamma_x\in W^{r,p}$.   For the forward implication, assume $\Gamma_x\in W^{r,p}$, $0\leq r\leq s$.   Then solving for the derivatives of $J$ on the right hand side of (\ref{connection_transfo}) yields
\beq \label{J_derivatives}
\tfrac{\partial }{\partial x^\rho}  J^\alpha_\nu =   J^\alpha_\mu (\Gamma_x)^\mu_{\rho\nu}
-  J^\beta_\rho J^\gamma_\nu  \, (\Gamma_y)^\alpha_{\beta \gamma}.
\eeq
Now first this implies that $J$ must be at least as regular as $\Gamma_x$.   That is, if $J$ were less regular than $\Gamma_x,$ then both terms on the right hand side of (\ref{J_derivatives}) would be at least as regular as $J$, and then (\ref{J_derivatives}) would imply that the full gradient $\nabla J$ on the left hand side of (\ref{J_derivatives}) was at least as regular as $J$, and this is a contradiction.   So it must be that $J$ is at least as regular as $\Gamma_x$.  It follows from the assumption $r\leq s$ that the right hand side of (\ref{J_derivatives}) is at least as regular as $\Gamma_x$, (the object of lowest regularity).   We conclude that the left hand side of (\ref{J_derivatives}) is at least as regular as $\Gamma_x$, i.e., $\nabla J\in W^{r,p}$, which implies $J\in W^{r+1,p}$.  This establishes the forward implication, and completes the proof of Lemma \ref{Lemma_BigStart}.
\QED

If $\Gamma_x$ and $\Gamma_y$ have the same regularity, $r=s$, then Lemma \ref{Lemma_BigStart} allows for the possibility that $J$ can have, by cancellation of terms on the right hand side of \eqref{J_derivatives}, any regularity above $W^{r+2,p}$, including $C^\infty$.  This is consistent with the fact that, by (\ref{connection_transfo}), high regularity transformations preserve the regularity of a connection. But if $\Gamma_x$ and $\Gamma_y$ have different levels of regularity, e.g. $\Gamma_x\notin W^{r',p}$ for any $r'>r$, then Lemma \ref{Lemma_BigStart} implies that the regularity of the Jacobian is locked in at precisely one derivative above the lower connection regularity.  This is essentially because the connection transformation law incorporates all the derivatives of $J$, so there are no possible cancellations in the transformation law which can change the regularity of $J$ relative to the regularity of $\Gamma_x$ when $r<s$ and $\Gamma_x$ is no more regular than $W^{r,p}$.     An important consequence of Lemma \ref{connection_transfo} is that transforming a $W^{s,p}$ connection by a coordinate transformation $y\to x$ of regularity below $W^{s+2,p}$ always results in a loss of connection regularity, i.e., singular transformations always create singular connections. Conversely, only coordinate transformations $x\to y$ of precisely the regularity $W^{r+2,p}$ hold the possibility of lifting the regularity of a $W^{r,p}$ connection to a higher regularity.   We record this as follows:

\begin{Corollary} \label{Cor_BigStart}
Assume $\Gamma_x\in W^{r,p}$ but $\Gamma_x\notin W^{r',p}$ for $r'>r$ on some open set $\Omega_y$. Then any coordinate transformation $x\to y$ which lifts the regularity of the components $\Gamma_x\in W^{r,p}$  to $\Gamma_y\in W^{s,p}$, $s>r$, must have precisely the regularity $W^{r+2,p}$, and not $W^{r'+2,p}$ for any $r'>r$.   
\end{Corollary}

\Proof 
Since $\Gamma_x\in W^{r,p}$, the forward implication in Lemma \ref{connection_transfo} implies $J\in W^{r+1,p}$, and hence the transformation $x\to y$ is $W^{r+2,p}$ regular.  Moreover, if $J\in W^{r'+1,p}$ for $r'>r$, then the backward implication in Lemma \ref{connection_transfo} implies $\Gamma\in W^{r',p}$ as well, contradicting our assumption that $\Gamma_x$ is no more regular than $W^{r,p}$.   Thus the transformation $x\to y$ has precisely the regularity $W^{r+2,p}$.
\QED

Corollary \ref{Cor_BigStart} puts a constraint on the regularity of an atlas within which coordinate transformations sufficient to lift the regularity of a connection can be found, if they exist. By the corollary, a regularizing atlas must have regularity $W^{r+2,p}$, and hence will always lie within the maximal $W^{r+2,p}$ extension of a given atlas, but not within any atlas smoother than $W^{r+2,p}$.  It follows that for $r\geq 0$, the largest relevant atlas within which all regularizing coordinate transformations must lie, if they exist, is the $W^{2,p}$ extension of any given atlas.

\section{The Local Theory of Essential Regularity}  \label{Sec_local}

Equipped with Lemma \ref{Lemma_BigStart} together with the one step local regularization guaranteed by the RT-equations in Theorems \ref{Thm_Smoothing_high} - \ref{Thm_Smoothing_low} above, we can now establish the following local theory of essential regularity, including local versions of Theorems \ref{Thm1} and Theorem \ref{Cor1}. In particular, this characterizes the {\it local} structure of apparent singularities, which we take to mean connections given in a coordinate system with components exhibiting a regularity below its essential regularity. To start we define the following local notion of essential regularity: 

\begin{Def}  \label{Def_ess-reg}
Let $\Gamma_x \in L^p(\Omega_x)$ denote the components of a connection $\Gamma$ given in some coordinate system $x$ defined in a neighborhood $\Omega$ of a point $P\in\mathcal{M}$.     Define ${\rm ess}_{P}(\Gamma)$, the essential regularity  of $\Gamma$ in a neighborhood of $P,$ to be the largest integer $m\geq 0$ such that there exists a $W^{2,p}$ coordinate transformation $x \to y$, defined on a neighborhood $\Omega'\subset\Omega$ of $P$, such that $\Gamma_y \in W^{m,p}$ in $\Omega'$, if $m<\infty$ exists.  We say that ${\rm ess}_{P}(\Gamma) = \infty$ if for every $s\geq 0$ there exist a coordinate system $y$ such that $\Gamma_y \in W^{s,p}$ in some neighborhood of $P$.\footnote{Note that ${\rm ess}_{P}(\Gamma) = \infty$ includes the case that the connection is in $C^\infty$ in some coordinate system, but it also includes the possibility that no fixed coordinate system exist in which the connection is in $C^\infty$ in a neighborhood of $P$.} 
\end{Def}

\noindent As before, we always use $0 \leq m \leq \infty$ to denote the essential regularity of a connection, $m \equiv {\rm ess}_{P}(\Gamma)$. 

We now establish a local version of Theorem \ref{Cor1} which characterizes essential regularity in terms of a hierarchy of regularities between the connection and its curvature. 

\begin{Thm} \label{Cor1-local}  Assume $\Gamma_x\in W^{s,p}(\Omega_x)$ in a neighborhood $\Omega$ of a point $P\in\mathcal{M}$ for $n<p<\infty$, $s\geq0$, (but $p>4$ in case $n\leq 3$ and $s=0$), and $n\geq 2$.  Then:  
\begin{itemize}[leftmargin=.23in]  \setlength\itemsep{.5em}      
\item[{\bf (1)}] ${\rm ess}_{P}(\Gamma)=s$ if and only if $\Riem(\Gamma_x)\in W^{s-1,p}(\Omega_x')$ in some neighborhood $\Omega'$ of $P$, and $\Riem(\Gamma_x) \not\in W^{s,p}$ in any neighborhood of $P$;
\item[{\bf (2)}] ${\rm ess}_{P}(\Gamma)=s+1$ if and only if $\Riem(\Gamma_x)\in W^{s,p}(\Omega_x')$ in some neighborhood $\Omega'\subset\Omega$ of $P$, and $\Riem(\Gamma_x) \not\in W^{s+1,p}$ in any neighborhood of $P$;
\item[{\bf (3)}]  ${\rm ess}_{P}(\Gamma)\geq s+2$ if and only if $\Riem(\Gamma_x)\in W^{s+1,p}(\Omega_x')$ in some neighborhood $\Omega'\subset\Omega$ of $P$.   
\end{itemize}
\end{Thm}

\Proof
The proof of Theorem \ref{Cor1-local} is based on the regularization of connections by the RT-equations in Corollary \ref{Thm_Smoothing_low-cor} and Theorem \ref{Thm_Smoothing_high}, in combination with the precise regularity control of Jacobians which alter the connection regularity in Lemma \ref{Lemma_BigStart}, and the asynchronicity between the transformation law of the connection and the curvature, the former involving Jacobian derivatives and the latter involving only undifferentiated Jacobians.  To begin, recall that the transformation law for components of the Riemann curvature tensor under coordinate transformation $y^\alpha\to x^i $, with Jacobians $J^\alpha_\mu \equiv \frac{\partial y^\alpha}{\partial x^\mu}$ and $(J^{-1})^\mu_\alpha \equiv \frac{\partial x^\mu}{\partial y^\alpha}$, is given by
\beq \label{Riemann_transfo}
\Riem(\Gamma_x)^\tau_{\mu \nu \rho}
= (J^{-1})^\tau_\delta \; J^\alpha_\mu J^\beta_\nu J^\gamma_\rho \; \,\Riem(\Gamma_y)^\delta_{\alpha \beta \gamma},
\eeq
while the transformation law for connections is given in \eqref{connection_transfo} as
\beq \label{conn-transfo}
(\Gamma_x)^\mu_{\rho\nu}
= (J^{-1})^\mu_\alpha \Big( J^\beta_\rho J^\gamma_\nu  \, (\Gamma_y)^\alpha_{\beta \gamma}   +  \tfrac{\partial }{\partial x^\rho}  J^\alpha_\nu \Big).
\eeq
Here $\Gamma_x\equiv (\Gamma_x)^i_{jk}$ denotes the components of $\Gamma$ in $x$-coordinates, and $\Riem(\Gamma_x) \equiv \Riem(\Gamma_x)^i_{jkl}$ denotes the components of the Riemann curvature of $\Gamma$ in $x$-coordinates, which we view as functions of $x$-coordinates defined on $\Omega_x \equiv x(\Omega) \subset \R^n$; and $(\Gamma_y)^\alpha_{\beta \gamma} $ and $\Riem(\Gamma_y)^\delta_{\alpha \beta \gamma}$ are defined analogously. Note that $\Gamma_x \in W^{s,p}(\Omega_x)$ implies directly that the curvature has at least regularity $\Riem(\Gamma_x)\in W^{s-1,p}(\Omega_x)$, by the defining formula $\Riem(\Gamma_x) = d\Gamma_x + \Gamma_x \wedge \Gamma_x$.  We now prove Cases (1) - (3) of Theorem \ref{Cor1-local} separately. For this we assume that the connection components $\Gamma_x$  arose from transforming $y\to x$ from a coordinate system $y$ in which the connection exhibits its essential regularity, $\Gamma_y=m= {\rm ess}_P(\Gamma)$. 

{\bf Case (1):} To prove the forward implication, assume $\Gamma_x \in W^{s,p}(\Omega_x)$ exhibits essential regularity, that is, $\ess_P(\Gamma)=s$. Assume for contradiction that $\Riem(\Gamma_x) \in W^{s,p}(\Omega_x')$ in some neighborhood $\Omega'\subset \Omega$ of $P$. By Corollary \ref{Thm_Smoothing_low-cor} (for $s=0$) and Theorem \ref{Thm_Smoothing_high} (for $s\geq 1$), the RT-equations would then yield the existence of a coordinate system $y$ on some neighborhood $\Omega''$ of $P$ such that $\Gamma_y \in W^{s+1,p}(\Omega_y'')$, in contradiction to our incoming assumption that $\ess_P(\Gamma)=s$.

To prove the backward implication, assume $\Riem(\Gamma_x)\in W^{s-1,p}(\Omega_x)$ together with $\Riem(\Gamma_x)~\not\in~W^{s,p}(\Omega_x')$ for any neighborhood $\Omega'\subset\Omega$ of $P$. Assume for contradiction that $\ess_P(\Gamma)\geq s+1$. This implies there exists coordinates $y$ on some neighborhood $\Omega''$ of $P$ such that $\Gamma_y \in W^{s+1,p}(\Omega_y'')$. By Lemma \ref{Lemma_BigStart} the Jacobian $J$ of the coordinate transformation $x\to y$ and its inverse $J^{-1}$ have regularity $W^{s+1,p}$, one derivative above the regularity of $\Gamma_x$. Moreover, from the formula $\Riem(\Gamma_y) = d\Gamma_y + \Gamma_y \wedge \Gamma_y$ defining the curvature, it follows that $\Riem(\Gamma_y)\in W^{s,p}(\Omega_y'')$. Thus transforming $\Riem(\Gamma_y)\in W^{s,p}(\Omega_y'')$ according to the transformation law \eqref{Riemann_transfo} with the $W^{s+1,p}$ Jacobians $J$ and $J^{-1}$ would maintain the curvature regularity and imply $\Riem(\Gamma_x) \in W^{s,p}(\Omega_x'')$, in contradiction to our incoming assumption.

{\bf Case (2):} For the forward implication assume ${\rm ess}_{P}(\Gamma)=s+1$. Assume now for contradiction that $\Riem(\Gamma_x) \in W^{s+1,p}(\Omega_x')$ in some neighborhood $\Omega'\subset \Omega$ of $P$. Applying first Corollary \ref{Thm_Smoothing_low-cor} and then Theorem \ref{Thm_Smoothing_high}  (if $s=0$), or applying Theorem \ref{Thm_Smoothing_high} twice (if $s\geq 1$), it follows that there exists a coordinate system $y$ on some neighborhood $\Omega''$ of $P$ such that $\Gamma_y \in W^{s+2,p}(\Omega_y'')$; (note that the $W^{s+1,p}$ curvature regularity is preserved under the one-step regularization by the RT-equation since the regularizing Jacobian is $W^{s+1,p}$ regular). This contradicts our incoming assumption that $\ess_P(\Gamma)=s+1$.

To prove the backward implication, assume $\Riem(\Gamma_x)\in W^{s,p}(\Omega_x)$ together with $\Riem(\Gamma_x)~\not\in~W^{s+1,p}(\Omega_x')$ for any neighborhood $\Omega'\subset\Omega$ of $P$. Assume for contradiction that $\ess_P(\Gamma)\geq s+2$. Then there exists coordinates $y$ on some neighborhood $\Omega''$ of $P$ such that $\Gamma_y \in W^{s+2,p}(\Omega_y'')$. By Lemma \ref{Lemma_BigStart} the Jacobian $J$ of the coordinate transformation $x\to y$ and its inverse $J^{-1}$ have regularity $W^{s+1,p}$, one derivative above $\Gamma_x$. Moreover, by $\Riem(\Gamma_y) = d\Gamma_y + \Gamma_y \wedge \Gamma_y$, it follows that $\Riem(\Gamma_y)\in W^{s+1,p}(\Omega_y'')$. Thus transforming $\Riem(\Gamma_y)\in W^{s+1,p}(\Omega_y'')$ by \eqref{Riemann_transfo} with the $W^{s+1,p}$ Jacobians $J$ and $J^{-1}$ would imply $\Riem(\Gamma_x) \in W^{s+1,p}(\Omega_x'')$, in contradiction to our incoming assumption. 

{\bf Case (3):} For the forward implication assume ${\rm ess}_{P}(\Gamma)\geq s+2$. This implies there exists coordinates $y$ defined on some neighborhood $\Omega$ of $P$ such that $\Gamma_y\in W^{s+2,p}(\Omega_y)$. By $\Riem(\Gamma_y) = d\Gamma_y + \Gamma_y \wedge \Gamma_y$, the curvature has regularity $\Riem(\Gamma_y) \in W^{s+1,p}(\Omega_y)$, and by Lemma \ref{Lemma_BigStart} the Jacobian $J$ of the transformation $x\to y$ and its inverse $J^{-1}$ have regularity $W^{s+1,p}$. Thus, transforming $\Riem(\Gamma_y) \in W^{s+1,p}(\Omega_y)$ to $x$-coordinates according to \eqref{Riemann_transfo} implies $\Riem(\Gamma_x)\in W^{s+1,p}(\Omega_x)$, as claimed.\footnote{Note that the curvature can have arbitrarily more regularity than the connection in coordinates where the connection is below its essential regularity. The basic example for this would be the Euclidean metric transformed by any low regularity transformation; the Riemann curvature would always be zero. What is not clear is whether the curvature can ever be more regular than one derivative below the essential regularity of the connection in coordinates where the connection is at least two derivatives below its (finite) essential regularity, cf. Case (3) of Theorem \ref{Cor1-local}.}  

For the backward implication assume that $\Riem(\Gamma_x)\in W^{s+1,p}(\Omega_x')$ on some neighborhood $\Omega'\subset\Omega$ of $P$. Applying first Corollary \ref{Thm_Smoothing_low-cor} and then Theorem \ref{Thm_Smoothing_high} (if $s=0$, otherwise apply Theorem \ref{Thm_Smoothing_high} twice), there exist a coordinate system $y$ on some neighborhood $\Omega''$ of $P$ such that $\Gamma_y \in W^{s+2,p}(\Omega_y'')$. The essential regularity of $\Gamma$ is thus $W^{s+2,p}$ or higher, that is, ${\rm ess}_{P}(\Gamma)\geq s+2$. This completes the proof.   
\QED

Case {\bf (3)} of Theorem \ref{Cor1-local} immediately applies to any connection $\Gamma$ with $ess_P(\Gamma)=\infty.$  We record this in the following Corollary: 

\begin{Corollary}\label{CorInfinity}
Assume $ess_P(\Gamma)=\infty$, and assume $\Gamma_x\in W^{s,p}(\Omega_x)$, for $2\leq n<p<\infty$, $s\geq0$, (but $p>4$ in case $n\leq 3$ and $s=0$), in a coordinate system $x$ defined on an open set $\Omega\subset\mathcal{M}$ of $P$. Then $\Riem(\Gamma_x)$ is at least $W^{s+1,p}$ regular in some neighborhood $\Omega'\subset\Omega$ of $P$.
\end{Corollary} 

Based on Theorem \ref{Cor1-local} it is now easy to understand how to produce a connection below its essential regularity. This is based on the following asymmetry between the connection transformation law \eqref{conn-transfo} and the tensor transformation law \eqref{Riemann_transfo}: \eqref{conn-transfo} involves Jacobian derivatives, while \eqref{Riemann_transfo} involves only the undifferentiated Jacobian. It then follows directly by Lemma \ref{Lemma_BigStart} that transforming a $W^{m,p}$ connection $\Gamma$ by a Jacobian of regularity below $W^{m,p}$ always produces a connection below its essential regularity. Conversely, to construct examples of connections with essential regularity $W^{m,p}$ locally, simply take any collection of $W^{m,p}$ functions $\Gamma$, (viewed as the components of a matrix valued differential form), such that their exterior derivative (and hence their Riemann curvature) is one derivative less regular than these functions.   

We are now ready to prove the following local version of Theorem \ref{Thm1}.

\begin{Thm}\label{Thm1-local}  
Assume $\Gamma_x\in W^{s,p}(\Omega_x)$ in some coordinate system $x$ defined on a neighborhood $\Omega$ of a point $P\in\mathcal{M}$,  for $n<p<\infty$, $s\geq0$, (but $p>4$ in case $n\leq 3$ and $s=0$), and $n\geq 2$.  
\begin{itemize}[leftmargin=.3in]  \setlength\itemsep{.5em} 
\item[{\bf (i)}] There exists a neighborhood $\Omega'\subset\Omega$ of $P$ and a coordinate transformation $x\to y$ of regularity $W^{s+2,p}$ on $\Omega'$, such that  $\Gamma_y\in W^{s+1,p}(\Omega'_y)$ if and only if $\Riem(\Gamma_x)\in W^{s,p}$ in some neighborhood of $P$. 
\item[{\bf (ii)}] If $\ess_{P}(\Gamma)<\infty$, then subsequent use of the RT-equations yields a $W^{s+2,p}$ coordinate transformation $x\to y$ defined on some open neighborhood $\Omega'$ of $P$ such that $\Gamma_y$ exhibits its essential regularity, $\Gamma_y\in W^{m,p}(\Omega_y')$, $m= \ess_{P}(\Gamma)$.
\end{itemize}               
\end{Thm}      

\Proof 
{\bf (i):} The backward implication of Theorem \ref{Thm1-local}, when $\Riem(\Gamma_x)$ is assumed to be in $W^{s,p}$ in some neighborhood, follows immediately from our existence theorems for the $RT$-equations,  Theorem \ref{Thm_Smoothing_high} for the case $s\geq 1$, and Corollary \ref{Thm_Smoothing_low-cor} for the lowest regularity case $s=0$.   For the forward implication, when the existence of a smoothing transformation $x\to y$ which lifts $\Gamma_x\in W^{s,p}$ to $\Gamma_y\in W^{s+1,p}$ in a neighborhood of $P$ is assumed, we need to prove that the original curvature $\Riem(\Gamma_x)$ is at least as regular as $\Gamma_x$ in this neighborhood. This follows from the tensor transformation law \eqref{Riemann_transfo_Sec3} for the curvature in the form of Lemma \ref{Lemma_BigStart}.  Namely, if $\Gamma_y\in W^{s+1,p}$, then $\Riem(\Gamma_y)\in W^{s,p}$ by $\Riem(\Gamma_y)=d\Gamma_y + \Gamma_y\wedge\Gamma_y$, and since by Lemma \ref{Lemma_BigStart} the Jacobian $J^{-1}$ of the mapping $y\to x$ has regularity $J^{-1}\in W^{s+1,p}$, it follows from the tensor transformation law for the curvature that $\Riem(\Gamma_x)$ is at least as regular as $J^{-1}$ and $\Riem(\Gamma_y)$, that is, $\Riem(\Gamma_x)\in W^{s,p}$, $s\geq0$. This completes the proof of part (i) of Theorem \ref{Thm1-local}.

{\bf (ii):} Assume $m \equiv \ess_{P}(\Gamma)<\infty$. For concreteness we assume $\Gamma_x \in L^p(\Omega_x)$ and $\Riem(\Gamma_x) \in L^p(\Omega_x)$, $p>\max \{n,4\}$. Clearly, if $m=0$, then $\Gamma_x$ already has essential regularity. If $m=1$, then by {\bf (1)} of Theorem \ref{Cor1-local} $\Riem(\Gamma_x)\in L^p$. The $L^p$-existence theorem for the RT-equations in the form of Corollary \ref{Thm_Smoothing_low-cor} then establishes the existence of a $W^{2,p}$ coordinate transformation $x\to y$ on some neighborhood of $P$ lifting $\Gamma_x \in L^p$ to $\Gamma_{y} \in W^{1,p}$. If $m=2$, then by {\bf (2)} of Theorem \ref{Cor1-local} $\Riem(\Gamma_x)\in W^{1,p}$. Corollary \ref{Thm_Smoothing_low-cor} then yields a $W^{2,p}$ coordinate transformation $x\to y'$ on some neighborhood $\Omega'$ of $p$ lifting $\Gamma_x \in L^p$ to $\Gamma_{y'} \in W^{1,p}$. Since the Jacobian of the transformation and its inverse are both in $W^{1,p}$, the transformation law for the curvature implies $\Riem(\Gamma_{y'})\in W^{1,p}$. Thus, since $\Gamma_{y'} \in W^{1,p}$ and $\Riem(\Gamma_{y'})\in W^{1,p}$, our $W^{1,p}$-existence result for the RT-equations, Theorem \ref{Thm_Smoothing_high}, applies and yields a $W^{3,p}$ coordinate transformation $y' \to y$ on some neighborhood $\Omega''\subset \Omega'$ lifting $\Gamma_{y'} \in W^{1,p}$ to $\Gamma_{y''} \in W^{2,p}$.   Finally, if $m\geq 2$, continued use of the regularization of connections by Theorem \ref{Thm_Smoothing_high}, and composing the resulting transformations, yields a coordinate transformation $x\to y$ furnishing the essential connection regularity in some neighborhood $\Omega'$ of $P$, $\Gamma_y\in W^{m,p}(\Omega')$, $m= \ess_{\mathcal{M}}(\Gamma)$. This consecutive use of Theorem \ref{Thm_Smoothing_high} is possible since, after each regularization step, Part (ii) and (iii) of Theorem \ref{Cor1-local} imply that the curvature is regularized by at least one derivative as well, until the connection is regularized to its essential regularity. 
\QED 

Since the RT-equations in general only yield regularizing transformation on sub-neighborhoods in each step of the regularization, one can only expect a finite but arbitrary regularity gain in the case when $\ess_P(\Gamma) = \infty$, as recorded in the following corollary of Theorem \ref{Thm1-local}, (a local version of Corollary \ref{Cor_RT->ess_reg}).

\begin{Corollary}
Let $\Gamma_x\in W^{s,p}(\Omega_x)$, for $2\leq n<p<\infty$, $s\geq0$, (but $p>4$ in case $n\leq 3$ and $s=0$), and assume $ess_{P}(\Gamma)=\infty$. Then for each integer $0\leq s' <\infty$ subsequent use of the RT-equations yields a coordinate transformation which lifts $\Gamma$ to regularity $W^{s',p}$ in some neighborhood of $P$.
\end{Corollary}

\section{The Global Theory of Essential Regularity} \label{Sec_global}

We now give the proofs of our main results in Section \ref{Sec_Results} based on the results proven in Sections \ref{Sec_Jacobian} and \ref{Sec_local}. Recall that a $W^{s,p}$ atlas $\A$ on a manifold $\M$ ($\M$ considered fixed) is a collection of coordinate charts $(x_\alpha,\Omega_\alpha)$, $\Omega_\alpha \subset \M$ an open set and $x_{\alpha}: \Omega_{\alpha}\to\mathbb{R}^n$ an invertible mapping, such that the union $\cup_{\alpha\in\Lambda}\Omega_\alpha$ covers $\M$, and such that the transition maps $x_\alpha \circ y_\beta^{-1}$ of two charts $(x_\alpha,\Omega_\alpha)$ and $(x_\beta,\Omega_\beta)$ have regularity $W^{s,p}$ whenever $\Omega_\alpha\cap \Omega_\beta \neq \emptyset$. Again, we assume $p>n$ and restrict to integer values $s\geq 0$. Recall that given an atlas $\A$ of regularity at least $W^{s,p}$, then its maximal $W^{s,p}$ extension $\A^{\text{max}(s)}$, (i.e., its maximal $W^{s,p}$ atlas), is defined as the collection of all coordinate charts which contain the original atlas $\A$, and have $W^{s,p}$ transition maps on their overlaps, cf. \cite{Hirsch}. The maximal extension of a given atlas is unique \cite{Hirsch}. The extension $\A^{\text{max}(s)}$ of $\A$ includes restrictions of charts in $\A$ to smaller domains, as well as all charts on $\M$ whose transition maps have the same regularity on overlaps, and charts modified by composition with invertible functions on open sets of $\R^n$ whose compositions with coordinate systems in $\A$ produce $W^{s,p}$ regular maps on the overlaps. For our purposes here, we may consider two atlases on a topological space $\M$ to define the same manifold if they have the same maximal $W^{2,p}$ extension. 

A connection $\Gamma$ is said to be globally in $W^{s,p}$ if $\Gamma_{x} \in W^{s,p}(\Omega_{x})$ for every coordinate chart $(x,\Omega_x)$ in an atlas $\A$, and in this case we write $\Gamma \in W^{s,p}_\A$. Atlas regularity $W^{s+2,p}$ is required to preserve connection regularity $W^{s,p}$, and when convenient we denote such an atlas by $\A_s$, cf. Lemma \ref{Lemma_BigStart}.  Note that if $\Gamma \in W^{s,p}_{\A_s}$ with respect to some atlas $\A_s$, then $\Gamma \in W^{s,p}_{\A_s^{\text max(s+2)}}$, because the maximal $W^{s+2,p}$-extension of $\A_s$ consists of all charts with transition maps of regularity $W^{s+2,p}$, which all preserve the $W^{s,p}$ regularity of $\Gamma$. The regularity of the components of a connection depend on the regularity of the atlas, and so to capture the essential (``best'') regularity at each $p>n$ (assumed fixed), we cast the problem within the maximal atlas at the lowest regularity required to potentially lift the regularity of any connection in $L^p$ to higher regularity, which in this paper is $\A^{\text{max}}\equiv \A^{\text{max}(2)}$.      We can now prove the results stated in Section \ref{Sec_Results} in order of their appearance.

\subsection*{Proof of Lemma \ref{Lemma_atlas-control}}
Lemma \ref{Lemma_atlas-control} asserts that, if $\Gamma \in W^{s,p}_\A$, then the atlas $\A$ has regularity $W^{s+2,p}$. This follows from Lemma \ref{J_derivatives} applied to each coordinate chart in $\A$. That is, Lemma \ref{J_derivatives} asserts that the Jacobian of a transition map $y\circ x^{-1}$ is always one derivative more regular than the connection components in $x$- and $y$-coordinates. Thus, since by definition $\Gamma \in W^{s,p}_\A$ means that $\Gamma_{x} \in W^{s,p}(\Omega_{x})$ for every coordinate chart $(x,\Omega_x)$ in $\A$, all its transition maps $y\circ x^{-1}$ have regularity $W^{s+2,p}$. This implies atlas $\A$ has regularity $W^{s+2,p}$, and proves Lemma \ref{Lemma_atlas-control}. \hfill $\Box$ \\

\noindent To prove Theorem \ref{Thm1}, we establish the following lemma.

\begin{Lemma}\label{Lemma_subatlases}
Let $\Gamma\in W^{s,p}_{\mathcal{A}_s}$, $p>n$, $s\geq0$. Assume there exists a subatlas $\mathcal{A}_{s+1}$ of the maximal $W^{2,p}$ atlas $\Amax_s$ of $\A_s$ such that $\Gamma\in W^{s+1,p}_{\mathcal{A}_{s+1}}$. Then $\mathcal{A}_{s+1}$ is contained in the maximal $W^{s+2,p}$ atlas $\A_s^{\text{max}(s+2)}$ of $\A_s$. Moreover, any atlas in which the connection exhibits $W^{s,p}$ regularity is contained in $\A_s^{\text{max}(s+2)}.$ 
\end{Lemma} 

\Proof
Write $\A_s = \big\{ (x_\alpha,\Omega_\alpha) \big\}_{\alpha \in \mathscr{A}}$ and $\A_{s+1} = \big\{ (y_\beta,\Omega_\beta) \big\}_{\beta \in \mathscr{B}}$, for index sets $\mathscr{A}$ and $\mathscr{B}$. By assumption $\Gamma_{x_\alpha} \in W^{s,p}(\Omega_\alpha)$ and $\Gamma_{y_\beta} \in W^{s+1,p}(\Omega_\beta)$. Thus, on non-empty overlaps $\Omega_\alpha \cap \Omega_\beta \neq \emptyset$, Lemma \ref{J_derivatives} implies that $y_\beta \circ x_\alpha^{-1} \in W^{s+2,p}$ and $x_\alpha \circ y_\beta^{-1} \in W^{s+2,p}$. Thus $\mathcal{A}_{s+1}$ is contained in $\A_s^{\text{max}(s+2)}$, as claimed.  Finally note that by Lemma \ref{J_derivatives} the transition maps between any two atlases in which the connection exhibits $W^{s,p}$ regularity are $W^{s+2,p}$ related. Hence both atlases are contained in $\A_s^{\text{max}(s+2)}$ by its maximality, implying uniqueness. 
\QED

\subsection*{Proof of Theorem \ref{Thm1} - Part (i)}
Assume $\Gamma\in W^{s,p}_{\mathcal{A}_s}$, for $p>n$, $s\geq0$, and $p>\max\{4,n\}$ if $s=0$. Part (i) of Theorem \ref{Thm1} asserts that $\Gamma\in W^{s+1,p}_{\mathcal{A}_{s+1}}$ in some subatlas $\mathcal{A}_{s+1}$ of the maximal $W^{2,p}$ atlas $\Amax_s$ of $\A_s$ if and only if $\Riem(\Gamma)\in W^{s,p}_{\mathcal{A}_s}$.  
To prove the forward implication of (i), assume there exist some sub-atlas $\mathcal{A}_{s+1}$ of $\Amax_s$ such that $\Gamma\in W^{s+1,p}_{\mathcal{A}_{s+1}}$. From the identity $\Riem(\Gamma_x)=d\Gamma_x + \Gamma_x \wedge \Gamma_x$ applied to the components of the curvature in each coordinate chart of $\A_{s+1}$, we conclude that $\Riem(\Gamma)\in W^{s,p}_{\A_{s+1}}$. Now, by Lemma \ref{Lemma_subatlases}, both $\A_{s}$ and $\A_{s+1}$ are contained in the maximal $W^{s+2,p}$ atlas $\A_s^{\text{max}(s+2)}$ of $\A_s$, and thus all transition maps between charts in $\A_{s}$ and $\A_{s+1}$ defined on overlaps have regularity $W^{s+2,p}$. This suffices to preserve the curvature regularity and implies $\Riem(\Gamma)\in W^{s,p}_{\A_{s}}$, as claimed. 

To prove the backward implication of (i), assume $\Riem(\Gamma)\in W^{s,p}_{\mathcal{A}_s}$. Let $P \in \M$ and let $(x,\Omega)$ be a coordinate chart in $\A_s$ such that $P\in \Omega$. Then $\Riem(\Gamma_x)\in W^{s,p}(\Omega_x)$, and the backward implication of (i) of Theorem \ref{Thm1-local} implies the existence of a coordinate transformation $x\to y$ in $W^{s+2,p}$ defined on an open neighborhood $\Omega' \subset \Omega$ of $P$, such that $\Gamma_y \in W^{s+1,p}(\Omega'_y)$. Since this coordinate transformation is in $W^{s+2,p}$, it follows that the chart $(y,\Omega')$ is contained in the maximal $W^{s+2,p}$ atlas of $\A_s$ and thus also in its maximal $W^{2,p}$ extension $\Amax_s$. Continuing this regularization procedure for every $P \in \M$ yields a covering of coordinate charts $\big\{(y_P,\Omega_P)\big\}_{P\in\M} \subset \Amax_s$ such that $\Gamma_{y_P} \in W^{s+1,p}(\Omega_{y_P})$ for every $P\in\M$. By default, $\big\{(y_P,\Omega_P)\big\}_{P\in\M} \subset \Amax_s$ defines a $W^{s+2,p}$ sub-atlas of $\Amax_s$ which preserves the $W^{s+1,p}$ regularity of $\Gamma$. Lemma \ref{Lemma_BigStart} thus implies that its transition maps are all $W^{s+3,p}$ regular, from which we conclude that $\A_{s+1} \equiv \big\{(y_P,\Omega_P)\big\}_{P\in\M}$ defines a $W^{s+3,p}$ sub-atlas of $\Amax_s$, as claimed.  \hfill $\Box$ \\

For the proof of Theorem \ref{Thm1} (ii), recall that by Definition \ref{Def_ess-reg} we say a connection $\Gamma$ defined on ($\mathcal{M},\mathcal{A}$) has global essential regularity $m=\ess_{\mathcal{M}}(\Gamma)\geq 0$, $m\in \mathbb{N}_0$, if there exists a subatlas $\mathcal{A}_m$ of the maximal $W^{2,p}$ atlas $\Amax$ of $\A$ such that $\Gamma\in W^{m,p}_{\mathcal{A}_m}$,  and there does not exist a subatlas $\mathcal{A}_s$ of $\Amax$ in which $\Gamma\in W^{s,p}_{\mathcal{A}_s}$ with $s\in \mathbb{N}$ and  $s>m$. 

\subsection*{Proof of Theorem \ref{Thm1} - Part (ii)}
Assume again $\Gamma\in W^{s,p}_{\mathcal{A}_s}$, for $p>n$, $s\geq0$, and $p>\max\{4,n\}$ if $s=0$. Part (ii) of Theorem \ref{Thm1} asserts that, assuming $ess_{\mathcal{M}}(\Gamma)<\infty$, subsequent use of the RT-equations provides an algorithm for constructing an atlas $\A_m \subset \Amax$ in which $\Gamma$ exhibits its essential regularity $m=ess_{\mathcal{M}}(\Gamma)$. To prove this, let $P \in \M$ and let $(x,\Omega)$ be a coordinate chart in $\A_s$ such that $P\in \Omega$. By Theorem \ref{Thm1-local}, there exists a coordinate transformation $x\to y$ in $W^{s+2,p}$ defined on an open neighborhood $\Omega' \subset \Omega$ of $P$, such that $\Gamma_y \in W^{m,p}(\Omega'_y)$. Applying this argument to every point $P \in \M$ yields a covering of coordinate charts $\big\{(y_P,\Omega_P)\big\}_{P\in\M} \subset \Amax_s$ such that $\Gamma_{y_P} \in W^{m,p}(\Omega_{y_P})$ for every $P\in\M$. This covering defines a $W^{s+2,p}$ sub-atlas of $\A_s^{\text{max}(s+2)}$ which preserves the $W^{m,p}$ regularity of $\Gamma$, and Lemma \ref{Lemma_BigStart} implies that its transition maps are all $W^{m+2,p}$ regular. Thus $\A_{m} \equiv \big\{(y_P,\Omega_P)\big\}_{P\in\M}$ defines a $W^{m+2,p}$ sub-atlas of $\A_s^{\text{max}(s+2)} \subset \Amax_s$, as claimed.\footnote{One may take this atlas to be the maximal $W^{m+2,p}$ atlas. The maximal extension of $\A_m$ has the same maximal $W^{2,p}$ extension as the original atlas $\A$, and thus defines the same manifold.} This completes the proof of Theorem \ref{Thm1}. \hfill $\Box$

\subsection*{Proof of Corollary \ref{Cor_RT->ess_reg}}
Assume $\Gamma \in W^{s,p}_\A$ for some $W^{2,p}$ atlas $\A$ on $\M$, $s\geq 0$, and assume $ess_{\mathcal{M}}(\Gamma)=\infty$. Then Corollary \ref{Cor_RT->ess_reg} asserts that for each integer $0\leq s'<\infty$ subsequent use of the RT-equations yields an atlas $\A_{s'} \subset \Amax$ in which $\Gamma$ exhibits regularity $W^{s',p}_{\A_{s'}}$. This follows directly from the above regularization argument underlying the proof of Theorem \ref{Thm1} Part (ii) applied to $m\equiv s'$. \hfill $\Box$

\subsection*{Proof of Theorem \ref{Cor1}}
Assume $\Gamma\in W^{s,p}_{\A}$ is given on ($\mathcal{M},\mathcal{A}$), for $p>n$, $s\geq0$, and $p>\max\{4,n\}$ if $s=0$, and assume $\A=\A^{\text{max}(s+2)}$ is maximal. Theorem \ref{Cor1}, Case (1) then asserts that $\ess_\M(\Gamma)=s$ if and only if $\Riem(\Gamma)\in W^{s-1,p}_{\A}$ and $\Riem(\Gamma) \not\in   W^{s,p}_{\mathcal{A}}$. 

To prove the forward implication, assume $\ess_\M(\Gamma)=s$. In the setting of Case (1), this implies that there exists at least one point $P\in \M$ such that $\ess_P(\Gamma) =s$; (the essential regularity at other points might be higher). Theorem \ref{Cor1-local} (1) then implies that $\Riem(\Gamma_x) \not\in W^{s,p}_{\Omega_x}$ for all coordinate charts $(x,\Omega) \in \A$ with $\Gamma_x \in W^{s,p}(\Omega_x)$ and $P\in \Omega$. Thus $\Riem(\Gamma) \not\in W^{s,p}_{\mathcal{A}}$, while $\Riem(\Gamma)\in W^{s-1,p}_{\A}$ follows from $\Gamma\in W^{s,p}_{\A}$, just as claimed.

To prove the backward implication, assume $\Riem(\Gamma)\in W^{s-1,p}_{\A}$ and $\Riem(\Gamma) \not\in W^{s,p}_{\A}$. Assume for contradiction $\ess_\M(\Gamma)\geq s+1$. This implies in particular that there exists a subatlas $\A_{s+1}$ of $\Amax$ such that $\Gamma\in W^{s+1,p}_{\A_{s+1}}$. Thus $\Riem(\Gamma)\in W^{s,p}_{\A_{s+1}}$ and, by Lemma \ref{Lemma_subatlases}, $\A_{s+1}$ is contained in $\A=\A^{\text{max}(s+2)}$. Thus all transition maps between $\A_{s+1}$ and $\A$ are $W^{s+2,p}$ regular and preserve the curvature regularity, implying $\Riem(\Gamma) \in W^{s,p}_{\A}$, which is a contradiction.

Theorem \ref{Cor1} (2) asserts that $\ess_\M(\Gamma)=s+1$ if and only if $\Riem(\Gamma)\in W^{s,p}_{\A}$ and $\Riem(\Gamma) \not\in W^{s+1,p}_{\A}$. To prove the forward implication assume $\ess_\M(\Gamma)=s+1$. This implies that there exists a point $P\in\M$ such that $\ess_P(\Gamma)=s+1$. Theorem \ref{Thm1-local} (2) then implies that $\Riem(\Gamma) \not\in W^{s+1,p}(\Omega_x)$ for all $(x,\Omega) \in \A$ with $\Gamma_x \in W^{s,p}(\Omega_x)$ and  $P\in \Omega$. Thus $\Riem(\Gamma) \not\in W^{s+1,p}_{\A}$, as claimed. Moreover, $\ess_\M(\Gamma)=s+1$ implies the existence of a subatlas $\A_{s+1}$ of $\Amax_s$ such that $\Gamma \in W^{s+1,p}_{\A_{s+1}}$, which directly implies $\Riem(\Gamma)\in W^{s,p}_{\A_{s+1}}$. By Lemma \ref{Lemma_subatlases}, it follows that $\A_{s+1} \subset \A=\A^{\text{max}(s+2)}$, and the resulting $W^{s+2,p}$ regularity of the transition maps between $\A_{s+1}$ and $\A$ implies that $\Riem(\Gamma)\in W^{s,p}_{\A}$, as claimed. 

To prove the backward implication, assume $\Riem(\Gamma)\in W^{s,p}_{\A}$ and $\Riem(\Gamma) \not\in W^{s+1,p}_{\A}$. By Theorem \ref{Thm1} (i), there exists some subatlas $\A_{s+1} \subset \Amax$ such that $\Gamma \in W^{s+1,p}_{\A_{s+1}}$, (from the proof we know that $\A_{s+1} \subset \A^{\text{max}(s+2)}$), which implies that $\ess_\M(\Gamma)\geq s+1$. Assume for contradiction $\ess_\M(\Gamma)\geq s+2$. This implies that there exists a subatlas $\A_{s+2}$ of $\Amax$ such that $\Gamma\in W^{s+2,p}_{\A_{s+2}}$, and thus $\Riem(\Gamma)\in W^{s+1,p}_{\A_{s+2}}$. By Lemma \ref{Lemma_subatlases}, $\A_{s+2}$ is contained in $\A=\A^{\text{max}(s+2)}$, and thus all transition maps between $\A_{s+2}$ and $\A$ are $W^{s+2,p}$ regular and preserve the curvature regularity, which implies $\Riem(\Gamma) \in W^{s+1,p}_{\A}$, a contradiction. 

Finally, Theorem \ref{Cor1} (3) asserts that $\ess_\M(\Gamma)\geq s+2$ if and only if $\Riem(\Gamma)\in W^{s+1,p}_{\mathcal{A}}$. To prove the forward implication, assume $\ess_\M(\Gamma)\geq s+2$. This implies there exists an atlas $\A_{s+2} \subset \Amax$ such that $\Gamma \in W^{s+2,p}_{\A_{s+2}}$ and $\Riem(\Gamma) \in W^{s+1,p}_{\A_{s+2}}$. By Lemma \ref{Lemma_subatlases}, $\A_{s+2}$ is contained in $\A=\A^{\text{max(s+2)}}$, and the $W^{s+2,p}$ transition maps between $\A_{s+2}$ and $\A$ preserve the curvature regularity to imply $\Riem(\Gamma)\in W^{s+1,p}_{\mathcal{A}}$.

To prove the backward implication, assume $\Riem(\Gamma)\in W^{s+1,p}_{\mathcal{A}}$. By Theorem \ref{Thm1} (i), there exists some subatlas $\A_{s+1} \subset \Amax$ such that $\Gamma \in W^{s+1,p}_{\A_{s+1}}$ and thus also $\Riem(\Gamma) \in W^{s,p}_{\A_{s+1}}$. By Lemma \ref{Lemma_subatlases}, $\A_{s+1} \subset \A^{\text{max}(s+2)}$, and the $W^{s+2,p}$ transition maps imply $\Riem(\Gamma) \in W^{s+1,p}_{\A_{s+1}}$. Applying again Theorem \ref{Thm1} (i) implies the existence of a subatlas $\A_{s+2}$ such that $\Gamma \in W^{s+2,p}_{\A_{s+2}}$. This implies $\ess_\M(\Gamma) \geq s+2$ and completes the proof of Theorem \ref{Cor1}. \hfill $\Box$

\section{Estimates for Two-Step Regularizations} \label{Sec_estimates}

In this section, we establish local and global estimates on Sobolev norms of connections under a one- and two-step regularization. The one-step regularization by the RT-equations provides estimates on the regularized connections and the regularizing coordinate transformations in terms of the $W^{s,p}$-norm of the curvature in the starting coordinates, cf. Theorems \ref{Thm_Smoothing_high} and \ref{Thm_Smoothing_low}. This can be extended to estimates of a two-step regularization when the curvature in the starting coordinates is one derivative more regular than the connection, (cf. Case (3) of Theorem \ref{Cor1}), by using the corresponding higher $W^{s+1,p}$-norm on the curvature in the starting coordinates, because the curvature regularity stays fixed under a one- or two-step regularization.  However, the RT-equations provide no estimates for the regularization of the curvature. Thus they do not provide estimates for the connection when lifted to essential regularity by more than two steps, because this would require applying the RT-equations to the implicitly regularized curvature. Hence estimates for multi-step regularizations (more than two) appear out of reach. 

The estimates for the two-step regularization in the local case is stated in Corollary \ref{Cor_est_local} below. This is extended to a global setting on compact manifolds in Corollaries \ref{Cor_est_global1} and \ref{Cor_est_global2} below. We assume throughout that $p>n\geq 2$, $s\geq 0$.

\begin{Corollary}   \label{Cor_est_local}      
Assume $\Gamma \in W^{s,p}(\Omega_x)$ in some coordinate chart $(x,\Omega)$, such that 
\beq \label{local_estimates_1}
\|\Gamma_x \|_{W^{s,p'}(\Omega_x)} + \|\Riem(\Gamma_x)\|_{W^{s+1,p}(\Omega_x)} < M,	
\eeq
for some constant $M>0$, where either $s\geq 1$ and $p'=p>n$, or $s=0$, $p>n$ and $p'=\infty$. Then for every point $P \in \Omega$ there exist a neighborhood $\Omega'$ of $P$ (independent of $\Gamma_x$) such that the RT-equations yield a $W^{s+2,p}$ coordinate transformation $x\to y$ on $\Omega_x'$ such that 
\beq  \label{local_estimates_2}
\|\Gamma_y \|_{W^{s+2,p}(\Omega_y')} < C(M),	
\eeq
where $C(M)>0$ is a constant depending only on $\Omega_x, s, n, p$ and $M$, but independent of $\Gamma_x$.
\end{Corollary}

\Proof
For concreteness assume first that $s=0$. Then \eqref{local_estimates_1} implies the incoming bound \eqref{bound_incoming_ass} of Theorem \ref{Thm_Smoothing_low}, which implies the existence of a coordinate transformation $x\to \tilde{y}$ on some open set $\tilde{\Omega}$ with Jacobian $\tilde{J} \in W^{1,p}(\tilde{\Omega}_{\tilde{y}})$ such that $\Gamma_{\tilde{y}} \in W^{1,p}(\tilde{\Omega}_{\tilde{y}})$ and such that    
\beq \label{estimates_proof_1} 
\|\Gamma_{\tilde{y}} \|_{W^{1,p}(\tilde{\Omega}_{\tilde{y}})} + \|J\|_{W^{1,p}(\tilde{\Omega}_x)}  + \|J^{-1}\|_{W^{1,p}(\tilde{\Omega}_x)}  \leq \tilde{C}(M), 
\eeq    
for some constant $\tilde{C}(M)>0$ depending only on $\Omega_x, p, n$ and $M$.   Now, combining the bound implied on the curvature in \eqref{local_estimates_1} for $s=0$ with the bound implied on the Jacobians in \eqref{estimates_proof_1} yields
\beq \label{estimates_proof_2}
\|\Riem(\Gamma_{{\tilde{y}}})\|_{W^{1,p}(\Omega_{\tilde{y}})} < \tilde{C}(M),	
\eeq
for some new constant $\tilde{C}(M)>0$ depending only on $\Omega_x, p, n$ and $M$. Estimate \eqref{estimates_proof_2} together with the bound implied by \eqref{estimates_proof_1} on $\Gamma_{\tilde{y}}$ give
\beq \label{estimates_proof_3} 
\|\Gamma_{\tilde{y}}\|_{W^{1,p}(\Omega_{\tilde{y}})} + \|\Riem(\Gamma_{\tilde{y}})\|_{W^{1,p}(\Omega_{\tilde{y}})} 
\leq 2 \tilde{C}(M).
\eeq
This is the incoming bound of Theorem \ref{Thm_Smoothing_high}, which then yields the existence of a $W^{3,p}$ coordinate transformation $\tilde{y} \to y$ on a neighborhood $\Omega'$ of $P$ with Jacobian $J \in W^{2,p}(\Omega'_{\tilde{y}})$, such that $\Gamma_y \in W^{2,p}(\Omega'_y)$ and such that estimate \eqref{local_estimates_2} holds for $s=0$. Clearly, the transformation $x\to y$ is $W^{2,p}$ regular.  The case for $s\geq 1$ is analogous, requiring only the consecutive use of Theorem \ref{Thm_Smoothing_high}. 
\QED

Before extending Corollary \ref{Cor_est_local} to a global setting, note that Sobolev regularity of a tensor or connection is invariant under sufficiently regular coordinate transformations, but the value of the Sobolev norm itself is not invariant, except in the special case of a Riemannian metric. We address this problem here by basing norms on the Euclidean metric in each coordinate chart over a compact manifold. That is, Corollary \ref{Cor_est_local} extends to global estimates, provided the incoming bound holds in a suitable uniform sense over suitably uniform domains, which we implement by assuming $\M$ to be a compact manifold.    For one-step regularizations we have the following result.

\begin{Corollary}   \label{Cor_est_global1}
Assume $\Gamma \in W^{s,p}_{\A_s}$ for some $W^{s+2,p}$ atlas $\A_s$ on a compact manifold $\M$ such that 
\beq \label{global_estimates_1}
\|\Gamma_x \|_{W^{s,p'}(\Omega_x)} + \|\Riem(\Gamma_x)\|_{W^{s,p}(\Omega_x)} < M, \ \ \ \ \ \forall \ (x,\Omega) \in \A_s,	
\eeq
for some constant $M>0$, where either $s\geq 1$ and $p' = p > n$, or $s=0$, $p>n$ and $p'=\infty$. Then by the RT-equations there exists a $W^{s+3,p}$ atlas $\A_{s+1} \subset \Amax_s$ such that $\Gamma \in W^{s+1,p}_{\A_{s+1}}$ and
\beq  \label{global_estimates_2}
\|\Gamma_y \|_{W^{s+1,p}(\Omega_y)} < C(M), \ \ \ \ \ \forall \ (y,\Omega) \in \A_{s+1},
\eeq
where $C(M)>0$ is a constant depending only on $s, n, p, \A_s$ and $M$, but independent of $\Gamma$.
\end{Corollary}

\Proof
Applying Theorems \ref{Thm_Smoothing_low} if $s=0$ and Theorem \ref{Thm_Smoothing_high} if $s\geq 1$ at each point $P\in \M$ yields a covering of $\M$ by coordinate charts $\big((y_P,\Omega'_P)\big)_{P\in\M}$ contained in $\Amax_s$, such that $\|\Gamma_{y_P} \|_{W^{s+1,p}(\Omega'_P)} < C(M,P)<\infty$ for some constant depending on $M$ and the starting coordinate neighborhood $\Omega_P$ of $P$ in $\A_s$ (as well as the fixed values $s, n, p$). By compactness of $\M$ there exist a finite sub-cover $\big((y_{P_j},\Omega'_{P_j})\big)_{j = 1,...,N}$, which we take to be the atlas $\A_{s+1}$. By Lemma \ref{Lemma_BigStart}, $\A_{s+1}$ has indeed regularity $W^{s+3,p}$. Moreover, defining $C(M)$ as the maximum of $C(M,P_j)$ over $j=1,...,N$, we obtain the sought after estimate \eqref{global_estimates_2}.  This completes the proof.
\QED

The estimates for the local two-step regularization in Corollary \ref{Cor_est_local} extends to the global setting of compact manifolds as follows:

\begin{Corollary}   \label{Cor_est_global2}
Assume $\Gamma \in W^{s,p}_{\A_s}$ for some $W^{s+2,p}$ atlas $\A_s$ on a compact manifold $\M$ such that  
\beq \label{global_estimates_3}
\|\Gamma_x \|_{W^{s,p'}(\Omega_x)} + \|\Riem(\Gamma_x)\|_{W^{s+1,p}(\Omega_x)} < M, \ \ \ \ \ \forall \ (x,\Omega) \in \A_s,	
\eeq
for some constant $M>0$, where either $s\geq 1$ and $p' = p > n$, or $s=0$, $p>n$ and $p'=\infty$. Then by the RT-equations there exists a $W^{s+4,p}$ atlas $\A_{s+2}\subset \Amax_s$ such that $\Gamma \in W^{s+2,p}_{\A_{s+2}}$ and
\beq  \label{global_estimates_4}
\|\Gamma_y \|_{W^{s+2,p}(\Omega_y)} < C(M), \ \ \ \ \ \forall \ (y,\Omega) \in \A_{s+2},	
\eeq
where $C(M)>0$ is a constant depending only on $s, n, p, \A_s$ and $M$, but independent of $\Gamma$.
\end{Corollary}

\Proof
The proof is analogous to the one of Corollary \ref{Cor_est_global1}: Applying Corollary \ref{Cor_est_local} at each point $P\in \M$ yields a covering of $\M$ in $\Amax_s$ by coordinate charts such that $\|\Gamma_{y_P} \|_{W^{s+2,p}(\Omega'_P)} < C(M,P)<\infty$. We then choose  $\A_{s+2}$ as a finite sub-cover. By Lemma \ref{Lemma_BigStart}, $\A_{s+2}$ is a $W^{s+3,p}$ atlas, and defining $C(M)$ as the maximum of $C(M,P_j)$ over $j=1,...,N$, estimate \eqref{global_estimates_4} follows. 
\QED

\section*{Funding}
M.R. was supported by the City University of Hong Kong (Start-up Grant 7200748 and CityU Strategic Grant 7005839) and by the Hong Kong Research Grants Council (ECS-Grant
21306524 and GRF-Grant 11303326).

\section*{Acknowledgement}
We thank Michael Kunzinger and Roland Steinbauer for directing us to reference \cite{Hirsch} on $C^\infty$ differentiable structures. We thank Robert McCann for pointing out to us references \cite{McCann_etal, CalabiHartman, MyersSteenrod, Taylor_isometries}.

\end{document}